\documentstyle[prd,aps,preprint,tighten,epsfig]{revtex}
\newcommand{\order}[1]{$\cal{O}$(#1)}
\newcommand{\Eq}[1]{Eq.~\ref{#1}}
\newcommand{\Fig}[1]{Figure~\ref{#1}}
\newcommand{\Sect}[1]{Section~\ref{#1}}
\begin{document}

\newcommand{\be}{\begin{equation}}
\newcommand{\ee}{\end{equation}}

\newcommand{\bd}{\begin{displaymath}}
\newcommand{\ed}{\end{displaymath}}

\newcommand{\bea}{\begin{eqnarray}}
\newcommand{\eea}{\end{eqnarray}}

\newcommand{\ba}[1]{\begin{array}{#1}}
\newcommand{\ea}{\end{array}} 

\newcommand{\ord}[1]{\mathcal{O}(#1)}

\newcommand{\Cite}[1]{\cite{#1}}


\newcommand{\dd}{\partial}


\def\lvec#1{\setbox0=\hbox{$#1$}
    \setbox1=\hbox{$\scriptstyle\leftarrow$}
    #1\kern-\wd0\smash{
    \raise\ht0\hbox{$\raise1pt\hbox{$\scriptstyle\leftarrow$}$}}
    \kern-\wd1\kern\wd0}
\def\rvec#1{\setbox0=\hbox{$#1$}
    \setbox1=\hbox{$\scriptstyle\rightarrow$}
    #1\kern-\wd0\smash{
    \raise\ht0\hbox{$\raise1pt\hbox{$\scriptstyle\rightarrow$}$}}
    \kern-\wd1\kern\wd0}


\newcommand{\mbf}[1]{\mathbf #1}


\newcommand{\levi}[1]{\epsilon_{#1}}
\newcommand{\leviup}[1]{\epsilon^{#1}}


\def\cancel#1#2{\ooalign{$\hfil#1\mkern1mu/\hfil$\crcr$#1#2$}}
\def\slash#1{\mathpalette\cancel{#1}}

\def\simge{
    \mathrel{\rlap{\raise 0.511ex
        \hbox{$>$}}{\lower 0.511ex \hbox{$\sim$}}}}
\def\simle{
    \mathrel{\rlap{\raise 0.511ex 
        \hbox{$<$}}{\lower 0.511ex \hbox{$\sim$}}}}

\preprint{CU-TP-995, BNL-HET-00/27, RBRC-149}

\title{Non-perturbative Renormalisation of Domain Wall 
Fermions: Quark Bilinears}

\author{        
T.~Blum$^a$,
N.~Christ$^b$,
C.~Cristian$^b$,
C.~Dawson$^c$,
G.~Fleming$^b$\footnote{new address:                                    
Department of Physics, Ohio State University, Columbus, OH 43210, USA},
G.~Liu$^b$,
R.~Mawhinney$^b$,
A.~Soni$^c$,
P.~Vranas$^d$,
M.~Wingate$^{a*}$,
L.~Wu$^b$,
Y.~Zhestkov$^b$}

\address{
$^a$RIKEN BNL Research Center,
Brookhaven National Laboratory,
Upton, NY 11973}

\address{
$^b$Physics Department,
Columbia University,
New York, NY 10027}

\address{
$^c$Physics Department,
Brookhaven National Laboratory,
Upton, NY 11973}


\address{
$^d$IBM T. J. Watson Research Center, 
Yorktown Heights, NY, 10598}

\date{February 7, 2001}

\maketitle

\begin{abstract}

We find the renormalisation coefficients of the quark field and the flavour
non-singlet fermion bilinear operators for the domain wall fermion action, in
the regularisation independent (RI) renormalisation scheme. Our results are
from a quenched simulation, on a $16^3 \times 32$ lattice, with $\beta=6.0$
and an extent in the fifth dimension of $16$.
We also discuss the expected effects of the residual chiral symmetry breaking
inherent in a domain wall fermion simulation with a finite fifth dimension, and
study the evidence for both explicit and spontaneous chiral symmetry breaking
effects in our numerical results.
We find that the relations between different renormalisation factors predicted
by chiral symmetry are, to a good approximation, satisfied by our results
and that systematic effects due to the (low energy) spontaneous chiral
symmetry breaking and zero-modes can be controlled.
Our results are compared against the perturbative predictions for
both their absolute value and renormalisation scale dependence.
\end{abstract}

\section{Introduction}
\label{sec:intro}

Renormalisation of lattice operators is an essential ingredient needed to
deduce physical results from numerical simulations. In contrast with the
determination of hadronic masses, physical matrix elements can be determined
only if the normalisation of the appropriate lattice operators can be related
to that of the corresponding continuum operators, conventionally specified 
perturbatively at
short distances.  In principle, lattice perturbation theory may be used to
establish this connection.  However, lattice perturbation theory converges
slowly and the expansion parameter, the square of the lattice coupling
evaluated at the lattice scale, $g(a)^2$, decreases only as an inverse power
of $\ln(a)$.  This makes systematic improvement of perturbative results
essentially impossible.  This convergence may be improved when, following
ideas from continuum perturbation theory
\cite{Brodsky:1983gc}, a renormalised or "boosted"
\cite{Lepage:1993xa} coupling rather than the bare coupling is used as an
expansion parameter. Even so, considerable arbitrariness remains, and in
general it is extremely difficult to go beyond one loop order in such
calculations. To overcome these difficulties, Martinelli ${\it et. al.}$
\cite{Martinelli:1995ty} have proposed a promising
non-perturbative renormalisation procedure. This method 
has been previously used to determine renormalisation
coefficients for various operators using the Wilson
\cite{Gimenez:1998ue,Donini:1999sf,Giusti:1998gx,Becirevic:1998yg} and
staggered actions \cite{Aoki:1998hc}. The purpose of this work is to study the
application of this technique to the renormalisation of the quark field and
flavour non-singlet fermion bilinear operators for the domain wall fermion
action.

Domain wall fermions \cite{Kaplan:1993sg,Shamir:1993zy,Narayanan:1995gw}
provide an action, that at the expense of introducing a fifth dimension, has a
low energy theory with excellent chiral properties while at the same time
preserving exact flavour symmetry. These good chiral properties lead to a
suppression of the possible dimension five terms in the long-distance
effective Lagrangian implying that domain wall fermions define a lattice
version of QCD which is off-shell improved to ${\cal O}(a^2)$.  As we will
see, these domain wall off-shell Greens functions show remarkably reduced
lattice artifacts.  A study of operator renormalisation coefficients for this
action is useful, both because these numbers are needed for use in practical
calculations of physical quantities \cite{Blum:2000kn} and because it provides
an excellent test of the chiral properties of the domain wall fermion action
in practical simulations. In fact, we find that domain wall fermions perform
quite well for non-perturbative renormalisation with negligible contributions
from explicit chiral symmetry breaking.  This finding is in good agreement
with recent work on the chiral limit of quenched QCD with domain wall fermions
\cite{Blum:2000kn,AliKhan:2000iv}.

Careful operator normalisation is especially important for the domain wall
fermion method.  As is reviewed in Section~\ref{sec:DWF_action}, the
interpolating field conventionally used to create and destroy the physical
modes is exactly localised in the fifth-dimension on the right and left walls.
Since the actual physical modes extend somewhat into the fifth dimension, the
overlap between the interpolating field and the physical modes will be smaller
than one.  This implies a wave function renormalisation factor $(Z_q)$ which
differs from one even in the case of free fields.  For the eigenvectors
corresponding to the smallest 19 Dirac eigenvalues examined in the quenched,
$\beta=6.0$ calculation of Ref.~\cite{Blum:2000kn}, this overlap typically
varies between 75 and 85\%.  Fortunately, the non-perturbative methods
employed here \cite{Martinelli:1995ty} precisely include these effects.

We begin in \Sect{sect:npr} with a brief summary of the main issues involved
in applying the non-perturbative renormalisation method.  In \Sect{sect:dwf},
we give the domain wall fermion action and discuss the Ward-Takahashi
identities it obeys. \Sect{sec:chi_break} builds on this base to constrain the
ways in which explicit chiral symmetry breaking terms may enter low energy
matrix elements calculated using domain wall fermions. In \Sect{sect:num} we
give the details of our lattice simulations. \Sect{sect:renprop} describes the
renormalisation of the quark propagator, and in \Sect{sect:bil} we introduce
the quark bilinears and compute their renormalisation on the lattice in the
regularisation independent scheme.  After removing expected non-perturbative
pole terms, we look for effects of explicit chiral symmetry breaking and find
that they are negligible. In \Sect{sec:zqward}, we avail ourselves of the axial
Ward-Takahashi identity again to compute the quark wave function
renormalisation from the conserved vector and partially conserved axial-vector
currents. In \Sect{sect:spect} we calculate the renormalisation of the
non-conserved, local axial current from a ratio of its hadronic matrix element to the
hadronic matrix element of the partially conserved axial current and find good
agreement with the results of \Sect{sect:bil}. In \Sect{sec:rg} we convert the
renormalisation coefficients to renormalisation group invariant quantities by
dividing out the renormalisation group running. In
\Sect{sect:zq_prop} we discuss the calculation of the quark wavefunction
renormalisation from the propagator.

After comparing our non-perturbative results with recent perturbative 
calculations in \Sect{sect:pert}, we end with our conclusions. The details
of the exact conventions and equations used for the perturbative running
and matching are relegated to appendices.

\section{Non-Perturbative Renormalisation}
\label{sect:npr}
In the following the method
of non-perturbative renormalisation introduced in 
Ref.~\cite{Martinelli:1995ty} will be studied. This method uses
a renormalisation scheme that is defined by a set of conditions that
mandate the renormalised values of the operators of interest 
between external quark states, in a fixed gauge, at large
virtualities. As such these conditions may be expressed in any
regularisation scheme (and so this scheme is known as the regularisation
independent (RI) scheme). In
particular this allows the renormalisation factors to be defined in
the lattice regularisation, opening the way for renormalisation
factors to be directly calculated in
numerical lattice simulations. 

While calculating renormalisation factors from
lattice simulations neatly avoids the need to perform
analytic calculations using lattice perturbation theory, which
are both challenging and poorly behaved, doing so introduces
several issues that must be considered:
\begin{itemize}
\item  Calculating the matrix elements of the operators of interest
between external quark states requires a fixed gauge to be used. This allows
for the appearance of Gribov copies, possibly obscuring the required comparison 
with continuum perturbation theory where only the trivial copy appears.  Earlier
studies\cite{Paciello:1994gs} of the size of Gribov noise in the calculation 
of a gauge invariant normalisation factor as a ratio of two gauge-variant
amplitudes suggest this may not be an important difficulty for the parameters 
used here.  However, in future work, this difficulty can be avoided by taking 
two steps:  i) Impose the regularisation invariant normalisation condition in 
a sufficiently small physical volume that non-perturbative effects are 
suppressed.  ii) Begin the Landau gauge fixing procedure from a configuration 
that is in a completely-fixed axial gauge.  Taking these two precautions will 
insure that any effects of Gribov copies will be similar to other 
non-perturbative effects and will vanish as the comparison with perturbation 
theory is done at weaker and weaker coupling.
\item Numerical simulations are performed with a finite lattice spacing. This
provides a natural condition,
\be
|p| \ll \frac{1}{a} \, ,
\ee
over the momenta range for which a direct extraction of continuum quantities
is possible.
\item As the renormalisation factors are determined in a non-perturbative
calculation the contributions of propagating mesons, and in
particular pseudo-Goldstone bosons, must be identified and removed.
These effects may be reduced by working at high momenta,
with a natural condition for the absence of significant deviations being
\be
\Lambda_{QCD} \ll |p| \, .
\ee
\end{itemize}
Taking the last two points together suggests that
this technique relies on the existence of a ``window'' of momenta,
\be
\Lambda_{QCD} \ll |p| \ll \frac{1}{a} \, ,
\ee
for which the predictions of continuum perturbation theory should correctly describe
the form of the lattice data. In practical simulations however, it has been
found that the effects of deviations due the violations of both these
inequalities must be taken into
account~\cite{Gimenez:1998ue,Gockeler:1998ye,Becirevic:1998yg}.  

Fortunately, near either edge of this window, the form of deviations from
perturbative behaviour may be predicted. In the case of momenta too low, the
initial corrections may be described by an expansion in terms of 
momentum-suppressed 
condensate terms by use of the operator product expansion (OPE). In
turn, the first corrections to continuum-like behaviour may be taken into
account in terms of an expansion in the lattice spacing, $a$.

Another trivial consequence of the restricted range of momenta available in
current lattice simulations is the need for many phenomenological calculations
to be composed of continuum perturbation theory calculations
at high scales, that are then run down to scales accessible on the lattice and
combined with the lattice result. As the majority of the existing calculations
for the continuum perturbative results use renormalisation schemes that may
only be defined when using dimensional regularisation (such as the
$\overline{MS}$ scheme), perturbative matching calculations between these
schemes and the ones that may be defined in the lattice regularisation need to
be performed.

\section{Domain Wall Fermions}
\label{sect:dwf}
In this section the domain wall fermion formulation, as used in our simulations, 
will be reviewed. 
\subsection{Action}
\label{sec:DWF_action}
The domain wall fermion (DWF) method is a promising new approach to lattice QCD
introduced in Ref.~\cite{Shamir:1993zy}, which, at the expense of introducing an
extra, discrete, non-gauge dimension, provides drastically improved chiral
properties at finite lattice spacing while preserving exact symmetry under
vectorial flavour rotations. This is achieved by using an action in the
fifth dimension that is asymmetric between the left-handed and right-handed
components of the fermion field. Denoting the fifth co-ordinate as $s$, with 
\be
s \in 0,\cdots, L_s-1 \, ,
\ee
the massless action may be written as
\bea
S_{{\rm fermion}}(m_f=0)&=&
-\sum_{x,s}
\overline{\Psi}_{x,s}
\left\{
-\gamma_\mu
 \frac{1}{2} \left( \nabla^{+}_{\mu} + \nabla^{-}_{\mu} \right)
\right.
\label{dwf:D_DWF}
\\ \nonumber
&& 
\left.
+ \left[ 
\frac{1}{2} \nabla^{-}_{\mu}\nabla^{+}_{\mu} + M_5
\right]
+ P_L \partial^{+}_5 - P_R \partial^{-}_5
\right\}
\Psi_{x,s}  \, ,      
\eea
with
\be
Z = \int [dU] [d\overline{\Psi} d\Psi] \exp \left( - S_{{\rm gauge}} - 
S_{{\rm fermion}}
\right) \, .
\label{dwf:act}
\ee
In Eqs.~\ref{dwf:D_DWF} and \ref{dwf:act} $\Psi_{x,s}$ is the fermionic
field, $U_\mu(x)$ is the gauge field and  
\be
S_{{\rm gauge}} = \beta \sum_{P} 
\left(
1 - \frac{1}{3} \mathrm{Re\,Tr} \left[ U_P \right] 
\right) \, ,
\ee
with $\beta = 6/g_0^2$ and $g_0$ is the bare lattice coupling.
The projectors for the left and right-handed spinors are defined as,
\bea
P_L &=& \frac{1}{2} \left( 1 - \gamma^{5} \right)
\\ \nonumber
P_R &=& \frac{1}{2} \left( 1 + \gamma^{5} \right) \, .
\eea
The notation $\nabla^{\pm}_{\mu}$ has been used to
denote the discrete forward/backward covariant derivatives:
\begin{eqnarray}
\nabla^{+}_{\mu} \psi_x
&=&
\left[
U_{\mu}(x) \psi_{x+\mu}
-\psi_x 
\right]
\\
\nabla^{-}_{\mu} \psi_x
&=&
\left[
\psi_x
-
U^{\dagger}_{\mu}(x-\mu) \psi_{x-\mu}
\right] \, ,
\end{eqnarray}
and $\partial^{\pm}_\mu$ represents the corresponding derivative
with no gauge term. For the case of the derivative in
the fifth dimension, $\partial^{\pm}_5$, 
the domain wall is implemented by giving the 
derivative hard boundaries. For example
a one-dimensional $\partial_5^+$ acting on a space with four points may be
written in matrix form as
\be
\partial_5^+
\left(
\begin{array}{c}
\Psi_{x,0} \\
\Psi_{x,1} \\
\Psi_{x,2} \\
\Psi_{x,3} \\
\end{array}
\right)
=
\left(
\begin{array}{cccc}
-1&1&0&0 \\
0&-1&1&0 \\
0&0&-1&1 \\
0&0&0&-1 \\
\end{array}
\right)
\left(
\begin{array}{c}
\Psi_{x,0} \\
\Psi_{x,1} \\
\Psi_{x,2} \\
\Psi_{x,3} \\
\end{array}
\right) \, .
\ee
It should be noted that the action in \Eq{dwf:D_DWF} is actually the hermitian
conjugate of the action proposed in Ref.~\cite{Shamir:1993zy}. This change was
made for practical reasons related to compatibility with the existing Wilson
operator implementation for the QCDSP machine.

In the free theory, for $0 < M_5 < 2$, the effect of this is to produce a
 spectrum with one light fermionic mode, with exact chiral symmetry in the
 $L_s \rightarrow \infty$ limit, and $16L_s-1$ heavy modes. The wavefunction
 of this light mode has its right-handed component concentrated on the wall at
 $s=L_s-1$ and its left-handed component on the wall at $s=0$.  This light
 fermion mode
 may be studied by introducing an interpolating operator 
 of the form
\cite{Furman:1995ky}
\bea
q_x &=& P_L \Psi_{x,0} + P_R \Psi_{x,L_s-1} 
\label{eq:4-d_fermions}
\\ \nonumber
\overline{q}_x &=& \overline{\Psi}_{x,0} P_R + \overline{\Psi}_{x,L_s-1} P_L \,.
\eea
The above considerations also naturally lead to the introduction of an
explicit mass term to the action of the form
\be
S_{{\rm fermion}}(m_f) = S_{{ \rm fermion}}(m_f=0) + \sum_{x} \, m_f \, \overline{q} q \, ,
\label{eq:realdwf}
\ee
where $m_f$ is the bare quark mass.
In the free case, this leads to a  spectrum with one light fermion of mass   
\be
M_5 \left( 2 - M_5 \right)\left( m_f + \left( 1 - M_5\right)^{L_s}
\right)\, .
\label{dwf:mfree}
\ee 
Note that in the $L_s \rightarrow \infty$ limit this is
proportional to $m_f$, while for finite $L_s$ there remains a 
residual mass, $m_{{\rm res}}$, that acts as an additive
renormalisation to $m_f$.

However, the properties of domain wall fermions in the presence of gauge
fields is a much more difficult question. In particular while the form of the
mass of the light mode is expected to be proportional to $m_f +
m_{\mathrm{res}}$, the dependence of $m_{\mathrm{res}}$ on $L_s$ must be
determined.  Perturbative calculations
\cite{Aoki:1999ky,Aoki:1998hi,Aoki:1998vv,Aoki:1997xg,Blum:1999xi} have shown
that the existence of the light mode is stable to small perturbations and that
this mode has all chiral symmetry breaking proportional to $m_f$ as
$L_s \rightarrow \infty$. These
studies also highlight several issues that must be considered when
undertaking numerical simulations: 
\begin{enumerate}
\item The dependence of $m_{{\rm res}}$ on $L_s$ may no longer be 
of the simple exponential form shown in \Eq{dwf:mfree}.
\item $M_5$ undergoes a strong additive renormalisation. 
This is understandable, as the five dimensional problem has no approximate
chiral symmetry to protect it.
\end{enumerate}

Indeed, extensive numerical studies in the quenched approximation
\cite{Blum:2000kn,AliKhan:2000iv} have shown that the $L_s$ dependence of $m_{\mathrm{res}}$ 
does not fit a single exponential in the range $L_s = 12 \rightarrow 48$ for
lattices with the same lattice spacing ($a = 0.520 {\rm GeV}^{-1}$) as the
results in this paper. For $L_s=16$, the value used in this work,
$m_{\mathrm{res}}$ was found to be $\sim 4{\rm MeV }$ in the $\overline{MS}$
scheme at $2 \mathrm{GeV}$ \cite{Blum:2000kn}.  The strong additive
renormalisation of $M_5$ requires that an input value be chosen numerically
so that a single light mode forms and that its decay in $L_s$ is as rapid as
possible. It has been found that for even coarser lattices than used here such
a choice can be made \cite{Blum:1998ud,Blum:2000kn}.

\subsection{Lattice Ward-Takahashi  Identities}
For the purpose of analysing the consequences of the symmetries
of the action, it is convenient to introduce an extended mass
term, $M$, with flavour structure such that the mass term reads
\be
\overline{q}_L M^{\dagger} q_R + \overline{q}_R M q_L \,,
\label{dwf:extended_mass}
\ee
and so the mass term is invariant 
under a transformation of the quark fields 
and the mass matrix $M$ of the form
\bea 
\nonumber
q_L &\rightarrow& U_L q_L  \\ \nonumber 
q_R &\rightarrow& U_R q_R  \\  
M &\rightarrow& U_R \, M \, U_L^{\dagger}.
\label{dwf:gen}
\eea
Following Ref.~\cite{Furman:1995ky}, on a finite lattice, 
an exact vector Ward-Takahashi identity may be derived by
considering transformations of the 5-dimensional fermion field, $\Psi$,
such that
\bea
\nonumber
\delta_V \Psi_{x,s} &=& i \epsilon_{x}^{a} T^{a} \Psi_{x,s} \\
\delta_V \overline{\Psi}_{x,s} 
&=&  - i \epsilon_{x}^{a} \overline{\Psi}_{x,s} T^{a} \, ,
\label{dwf:vecgen}
\eea
where $\left\{T^a\right\}$ is the set of hermitian traceless matrices acting
on $SU(N_f)$ flavour-space. This leads to an exact Ward-Takahashi identity 
that reads:
\bea
\label{dwf:evec}
- \partial_\mu^{-} \langle {\cal V}_{\mu}^{a}(z) O(x_1,\cdots,x_n) \rangle
&&
\\ \nonumber
+ \langle \overline{q} \left[ M, T^{a}\right] q(z) \, O(x_1,\cdots,x_n) \rangle
&=&
-i\langle \delta^a O(x_1,\cdots,x_n) \rangle \, ,
\eea
where
\bea
\label{dwf:cvec}
{\cal V}_\mu^{a}(x)  &=&
 \frac{1}{2} \sum_s 
\left( 
\overline{\Psi}_{x+\mu,s} \left( 1+\gamma_\mu \right) 
U^{\dagger}_{x,\mu} T^{a}
\Psi_{x,s}
\right. 
\\\nonumber
&& \left. - 
\overline{\Psi}_{x,s} \left( 1 - \gamma_\mu \right) 
U_{x,\mu} T^{a}
\Psi_{x+\mu,s} 
\right)
\, .
\eea
For the case of axial transformations the analogous choice is a transformation
of the form:
\bea
\nonumber
\delta_A \Psi_{x,s} &=& i \epsilon_{x,s}^{a} T^{a} \Psi_{x,s} \\
\delta_A \overline{\Psi}_{x,s} 
&=& - i \epsilon_{x,s}^{a} \overline{\Psi}_{x,s} T^{a} \, ,
\label{dwf:axgen}
\eea
with
\be
\epsilon_{x,s}^a  
= 
\left\{
\begin{array}{rc}
\epsilon_x^a & 0 \leq s < L_s/2 \\
 - \epsilon_x^a & L_s/2 \leq s < L_s \, .
\end{array}
\right.
\ee
This leads to a Ward-Takahashi  identity of the form 
\bea
- \partial_\mu^{-} \langle {\cal A}_\mu^a(z) O(x_1,\cdots,x_n) \rangle
\label{dwf:eveca}
\\ \nonumber
+ \langle \overline{q} \left\{ M, T^{a}\right\}\gamma_5 q(z) \, O(x_1,\cdots,x_n) \rangle
\\ \nonumber
+
2 \langle J_{5q}^a \, O(x_1,\cdots,x_n) \rangle
&=&
- i \langle \delta^a_{A} O(x_1,\cdots,x_n) \rangle \, ,
\eea
where
\bea
\label{dwf:acur}
{ \cal A}^a_\mu &=& 
 \frac{1}{2} \sum_s 
\mathrm{sign}\left(
s- \frac{L_s-1}{2} 
\right)
\left( 
\overline{\Psi}_{x+\mu,s} \left( 1+\gamma_\mu \right) 
U^{\dagger}_{x,\mu} T^{a}
\Psi_{x,s}
\right. 
\\\nonumber
&& \left. - 
\overline{\Psi}_{x,s} \left( 1 - \gamma_\mu \right) 
U_{x,\mu} T^{a}
\Psi_{x+\mu,s} 
\right)
\\
J_{5q}^a  &=&
- 
\overline{\Psi}_{x,L_s/2-1}
P_L T^{a} \Psi_{x,L_s/2}
+ 
\overline{\Psi}_{x,L_s/2}
P_R T^{a} \Psi_{x,L_s/2-1}
\, .
\eea
Therefore, in contrast to the previous case, the axial current is not exactly
conserved. This is necessary both to provide a mechanism for physical terms due to
the $U(1)_A$ axial anomaly to enter the calculated amplitudes and also to allow
for explicit chiral symmetry breaking contributions at finite $L_s$. The
situation is analogous to that for Wilson fermions \cite{Bochicchio:1985xa},
where the role of $J_{5q}^a$ is played by the chiral variation of the Wilson
term, except that the contributions from $J_{5q}^a$ are expected to tend to
zero as $L_s\rightarrow \infty$ in the present case \cite{Furman:1995ky}. The
form of the contributions from $J_{5q}^a$ will be further discussed in the
next section.

\section{Operator Mixing  and Chiral Symmetry}
\label{sec:chi_break}
The major attraction of the domain wall fermion formalism is its ability to
decrease the size of chiral symmetry breaking by increasing the parameter
$L_s$, the distance between the two four-dimensional lattice boundaries to
which the left and right chiral modes are bound.  However, it is often
impractical or inefficient to choose such a large value of $L_s$ that all
chiral symmetry breaking effects from mixing between these walls can be
neglected. Thus, it is important to characterise the effects of this chiral
symmetry breaking and in this section we will determine how it can effect the
low energy physics of lattice QCD.  As we will see, this can be done as either
an expansion in the size of the wall-mixing effects, which for simplicity we
will denote by ${\cal O}(e^{-\alpha L_s})$ although the exact $L_s$ dependence
may be different, and/or as an expansion in the lattice spacing $a$.

This analysis is easily made by starting with the interpretation of chiral
symmetry proposed by Furman and Shamir \cite{Furman:1995ky}.  Here one
identifies the full $SU(N_f)_L\otimes SU(N_f)_R$ chiral symmetry of the continuum
theory as the independent $SU(N_f)$ rotation of the
fermion fields defined on the left- and right-hand
halves of the five-dimensional lattice:
\be
\begin{array}{rcll}
\Psi(x,s) &\rightarrow& U_L \Psi(x,s)  & \;\; 0 \le s \le L_s/2 -1 \\
\Psi(x,s) &\rightarrow& U_R \Psi(x,s) & \;\; L_s/2 \le s \le  L_s-1\,,\\
\end{array}
\label{eq:5-d_chiral}
\ee
where $U_L$ and $U_R$ are $N_f \times N_f$ special unitary matrices
belonging to the left and right factors of $SU(N_f)_L\otimes SU(N_f)_R$.
From \Eq{eq:4-d_fermions} it is clear that this
transformation will act on the four-dimensional quark fields as a standard
element of the full chiral symmetry.

Of course, the transformation in \Eq{eq:5-d_chiral}, whose generators are
given in \Eq{dwf:axgen} and \Eq{dwf:vecgen}, cannot be an exact symmetry of the five-dimensional
theory as the derivative terms in the fifth dimension, taken collectively,
couple the left and right hand walls and prevent such independent rotations of
this single, five-dimensional field.  However, in the low energy sector of the
theory this symmetry can be quite good.  The physical, chiral modes which
survive at low energy are expected to be exponentially bound to the walls with
an overlap that is suppressed as $L_s$ increases.  The higher energy modes
which can propagate freely between the walls are all far off-shell with
propagators which are necessarily also exponentially suppressed at long
distances, especially for the large distance $L_s$.

In order to characterise the effects of this controlled symmetry breaking that
comes from communication between the walls, we will generalise somewhat the
Dirac domain wall fermion operator of Shamir given in \Eq{eq:realdwf}. We will
introduce a special-unitary, flavour matrix $\Omega$ in the derivative term joining the
four-dimensional planes $s=L_s/2-1$ and $s=L_s/2$.  Thus, we will modify
\Eq{dwf:D_DWF} by adding the term:
\be
S_{\Omega} = -
\sum_x
\left\{
\overline{\Psi}_{x,L_s/2-1} P_L 
\left(
\Omega^\dagger -1
\right)
 \Psi_{x,L_s/2}
+
\overline{\Psi}_{x,L_s/2} P_R 
\left(
\Omega
-1
\right) \Psi_{x,L_s/2-1}
\right\} \, .       
\label{eq:omega}
\ee
If we include the transformation of the matrix $\Omega$,
\begin{equation}
\Omega \rightarrow U_R \, \Omega \, U_L^{\dagger}
\label{eq:Omega_trans} \, ,
\end{equation}
this generalised domain wall Dirac operator will now possess exact chiral
symmetry. Note, a comparison with \Eq{dwf:gen} shows that $\Omega$ 
transforms ``like a mass term''.

Thus, if we examine this generalised theory that includes the chiral 
matrix $\Omega$, all amplitudes will become functions of $\Omega$ but 
will exactly obey the chiral symmetry described by 
\Eq{eq:5-d_chiral} and \Eq{eq:Omega_trans}. Therefore, we need 
only understand how the matrix $\Omega$ will enter the low energy 
Green's functions of interest to determine in a precise way the 
transformation properties of the chiral symmetry breaking induced 
by mixing between the walls.

To zeroth order in $e^{-\alpha L_s}$, the fermion degrees of freedom 
will remain bound to the walls and propagation from one wall to the 
other can be neglected.  In such circumstances, the matrix $\Omega$
which is introduced at a point mid-way between the walls cannot enter,
the amplitude will be independent of $\Omega$ and hence naively invariant
under the full $SU(N_f)_L \otimes SU(N_f)_R$ chiral symmetry.  To the next
order, $\propto e^{-\alpha L_s}$, we expect phenomena which involve a
single propagation between $s=0$ and $s=L_s-1$.  Thus, the matrix 
$\Omega$ should enter linearly in such amplitudes.

An important application of this analysis is to constrain the form
of the effective continuum action which gives amplitudes that agree
with those of the domain wall theory through a given order
in the lattice spacing.  To leading order in the lattice spacing,
this effective Lagrangian has the standard continuum form.  The above 
analysis requires that the mass term in this leading order effective
Lagrangian must have the form:
\begin{equation}
Z_m m_f \overline{\psi}\psi 
+ 
c \left\{
\overline{\psi} \Omega^\dagger P_R \psi
+ 
\overline{\psi} \Omega P_L \psi
\right\}\, ,
\label{chiral:mass}
\end{equation}
where $c$ is a constant with the dimensions of mass.  Here the field $\psi$ 
represents a conventional
continuum multiplet of quark fields and all quantities carry their physical
dimensions.  The first piece is the normal chiral symmetry breaking introduced
by the input mass $m_f$.  The second comes from mixing between the walls and
is required by the extended symmetry
of \Eq{eq:5-d_chiral} and \Eq{eq:Omega_trans} to be linear in $\Omega$.  
Thus, this induced mass term can occur to first order in the mixing between 
the walls, permitting $c \propto e^{-\alpha L_s}/a$.  With the conventional
choice of Shamir, $\Omega_{a,b} = \delta_{a,b}$, the second term in
\Eq{chiral:mass} reduces to our usual residual mass term with $am_{\rm res}
\approx 10^{-3}$ \cite{Blum:2000kn}.

In a similar fashion the ${\cal O}(a)$ effective Lagrangian will
contain a clover term induced by mixing between the walls, again to 
first order in $e^{-\alpha L_S}$, since it also has the permitted 
$SU(N_f)_L \otimes SU(N_f)_R$ chiral structure:
\be
a c_1 \, \left\{
\overline{\psi} \sigma_{\mu \nu } F_{\mu\nu}\Omega^\dagger P_R \psi
+ 
\overline{\psi} \sigma_{\mu \nu } F_{\mu\nu}\Omega P_L \psi
\right\}.
\ee
where $c_1 \propto e^{-\alpha L_s}$ is \order{$a m_{\mathrm{res}}$}.  
Thus, such a term is suppressed both by the lattice spacing and by the 
smallness of $m_{\mathrm{res}}$.

If we extend these considerations to 
${\cal O}(a^2)$ terms in the effective Lagrangian, we can conclude 
that a four-fermi operator of the form
\begin{equation}
{c_2 a^2}(\overline{\psi}\psi)(\overline{\psi}\psi)\, ,
\end{equation}
where $c_2$ is a constant,
cannot occur to order $e^{-\alpha L_s}$. Since this operator will 
become a chiral singlet only when contracted with two powers of the 
matrix $\Omega$ or one power of $\Omega$ and one powers of the mass 
matrix $M$ of \Eq{dwf:extended_mass}, the coefficient of such an 
operator must contain a double suppression 
$c_2 \propto e^{- 2 \alpha L_s}$ or a further factor of $m_f$.

\section{Simulation Details}
\label{sect:num}
In the following discussions much use will be made of the momentum space quark
propagator in Landau gauge. The first step in calculating this quantity
is to fix the gauge. We implement Landau gauge fixing by iteratively sweeping over
all lattice sites, maximising the functional
\be
 \sum_{x,\mu} \left[ {\rm Tr} \left( U_\mu(x) + U_\mu^\dagger(x)
   \right) \right] \, .
\label{eq:gf}
\ee
At each lattice site we determine a gauge transformation matrix,
$g(x)$, which increases the value of \Eq{eq:gf}. The maximum is achieved
when
\be
  \sum_x {\rm Tr} \left[ B(x) B^\dagger(x) \right]
\label{eq:cond}
\ee
is zero, where
\be
  B(x) = A(x) - A^\dagger(x) - \frac{1}{3} {\rm Tr}(A - A^\dagger)
\ee
\be
    A(x) = \sum_\mu \left[ \tilde{U}_\mu^\dagger(x) +
           \tilde{U}_\mu(x-\mu) \right] 
\ee
and
\be
    \tilde{U}_\mu(x) = g(x) U_\mu(x) g^{\dagger}(x+\mu) \, .
\ee
In practice we stop when the quantity in \Eq{eq:cond} is smaller than $10^{-8}$.

On this gauge-fixed configuration the quark
propagator, $S\left(x,0\right)$, from one source, denoted as $0$, to all possible sinks is then calculated.  A discrete Fourier
transform is then performed over the sink positions giving,
\be
S\left(p,0\right) = 
\sum_x \exp \left( -i p^{{\rm latt}}\cdot x \right) S \left( x, 0 \right) \, ,
\ee
with
\be
p^{\mathrm{latt}}_\mu = \frac{2 \pi}{L_\mu}n_\mu \, ,
\label{num:mom}
\ee
where $\mu$ is one of $x$,$y$,$z$ or $t$ and $n_\mu$ may in principle
lie in the range $ 0 \rightarrow L_\mu -1 $. In practice, however, only a
subset of this range is used. 

Unless otherwise stated all the data that will be presented is from
calculations on a $16^3 \times 32 \times 16$ lattice (where the last number
refers to the extent of the lattice in the fifth dimension). The simulation
was performed at $\beta = 6.0$ with $2000$ 
heatbath sweeps between every configuration and
with $2000$ thermalisation sweeps performed at the outset. In total
$142$ configurations were generated.  
For this lattice size the momentum
range was restricted to those momenta for which $n_\mu = 0,1,2$ for $\mu =
x,y,z$ and $n_t = 0,1,2,3$. Quark propagators for $5$ bare masses,
$m_f=0.01$, 0.02, 0.03, 0.04 and 0.05 were calculated all using $M_5 =
1.8$.

The results will often be quoted against the square of the absolute
momentum, where this refers to the Euclidean inner product
of the momentum defined in Eq.~\ref{num:mom}. To be more specific
\be
\left( ap\right)^2 = \sum_{\mu} p^{\mathrm{latt}}_\mu \, p^{\mathrm{latt}}_\mu
\, ,
\ee
where $p$ is dimensionful.\section{The Renormalised Propagator}
\label{sect:renprop}
Before we treat the more complicated situation of the fermion
bilinears it is necessary to first consider the renormalisation of the quark
propagator. Neglecting, for the moment, potential contributions from lattice
artifacts the renormalised quark field may be defined as
\be
q_{\mathrm{ren}}(x)  = Z_q^{\frac{1}{2}} q_0(x) \, .
\label{qren}
\ee
If we similarly introduce a renormalised mass, defined
by
\be
m_{\mathrm{ren}} = Z_m m_0 \, ,
\ee
where $m_0$ represents a generic multiplicatively renormalised bare mass,
then the  renormalised propagator may  be written
\be
S_{\mathrm{ren}}\left( p , m_{\mathrm{ren}} \right) 
=
Z_q S_0 \left( p , m_0 \right) |_{m_0 = m_{\mathrm{ren}}/Z_m } \, . 
\ee
Both $Z_q$ and $Z_m$ are fixed in the RI scheme by requiring that 
the renormalised propagator obey the Euclidean space relations 
\be
\lim_{m_{\mathrm{ren}}\rightarrow 0}  - \frac{i}{12} \mathrm{Tr}
\left(
\frac{ \partial S^{-1}_{\mathrm{ren}} }
{ \partial \slash{p} } (p)
\right)_{p^2  = \mu^2} = 1
\label{zq:ward}
\ee
\be
\lim_{m_{\mathrm{ren}} \rightarrow 0} \frac{1}{12 \,m_{\mathrm{ren}}} \mathrm{Tr}\left(
S^{-1}_{\mathrm{ren}}(p)
\right)_{p^2  = \mu^2 }
=
1 \, .
\label{rimass}
\ee

While \Eq{zq:ward} and \Eq{rimass} seem to give a simple and
appealing way to calculate $Z_q$ and $Z_m$ by directly applying
them to the lattice propagators, the effect of both lattice artifacts
and spontaneous chiral symmetry breaking must be considered.

Lattice actions with explicit chiral symmetry breaking require
an additive renormalisation of the input mass, which
may be taken into account for domain wall fermions by making the replacement
\be
m_0 \rightarrow m_f + m_{\mathrm{res}} \, ,
\ee
in the equations above.  
However, the effects of lattice artifacts on the correct definition of the
renormalised and improved quark field are more complicated. They have been studied in
Ref.~\cite{Becirevic:1999kb} for Wilson fermions, where it is 
noted that there are three terms that may mix with the definition at \order{a}
in the lattice spacing, giving rise to an expression for the improved and
renormalised quark field of
\be
q_{\mathrm{ren}}  = Z_q^{\frac{1}{2}} 
\left(
1 + b_q ma
\right)
\left\{
1 + a{c'}_q \left( \slash{D}+ m_{\mathrm{ren}} \right) + a c_{{\rm NGI}} \slash{\partial}
\right\} 
q_0 \, ,
\label{zq:break}
\ee
where $\slash{\partial}$ may appear because the gauge is fixed.
If such extra terms appear then conditions must be found 
that allow them to be subtracted from the bare quark
field before \Eq{zq:ward} and
\Eq{rimass} may be applied.
In the context of simulations using \order{a} improved Wilson action at
$\beta=6.0$ these terms have been found to give significant contributions to
the form of the propagator \cite{Becirevic:1999kb}.  In particular $c_q'$ was 
found to be large. However, they all break
chiral symmetry and so, following the arguments of
\Sect{sec:chi_break}, should be suppressed by a factor of $O(am_{{\rm res}})$
for simulations using domain wall fermions.  As such, studying the form
of the propagator provides an excellent test of the chiral properties
of domain wall fermions.

The effects of spontaneous chiral symmetry breaking on the form of the
propagator are well known \cite{Politzer:1976tv,Pascual:1982jr}.  The most
noticeable effect is that the trace of the inverse propagator picks up an
extra contribution, which at lowest order in a power expansion in $1/p^2$ may
be described as
\be
\frac{1}{12}{\rm Tr} \left( S^{-1}_{\mathrm{ren}}(p) \right) = m_{\mathrm{ren}} + C_1 \frac{\langle
\overline{q} q \rangle}{p^2}
+ \cdots
\label{zq:ss}
\ee
where, at first order in perturbation theory, $C_1 = 4 \pi \alpha_s / 3 $. Putting
\Eq{zq:ss} and \Eq{zq:break} together, the predicted form for the
trace of the lattice quark propagator is
\bea
\frac{1}{12} \mathrm{Tr} \left( S_{latt}^{-1}(ap) \right) 
&=& 
\cdots +
\frac{a^3\langle \overline{q} q \rangle}{(ap)^2} C_1 Z_q
\label{zq:trin}
\\ \nonumber
&&+\, Z_m Z_q \left\{ am_f + am_{{\rm res}} \right\}
\\ \nonumber
&&+\,  2 \left( c_{{\rm NGI}}Z_q - {c'}_q  \right) (ap)^2 \,
\, +
\cdots \, ,
\eea 
where terms of \order{$m c_{{\rm NGI}}$} have been neglected.  

\begin{figure}[!htb]
\begin{center}
\epsfxsize=\hsize
\mbox{\epsfig{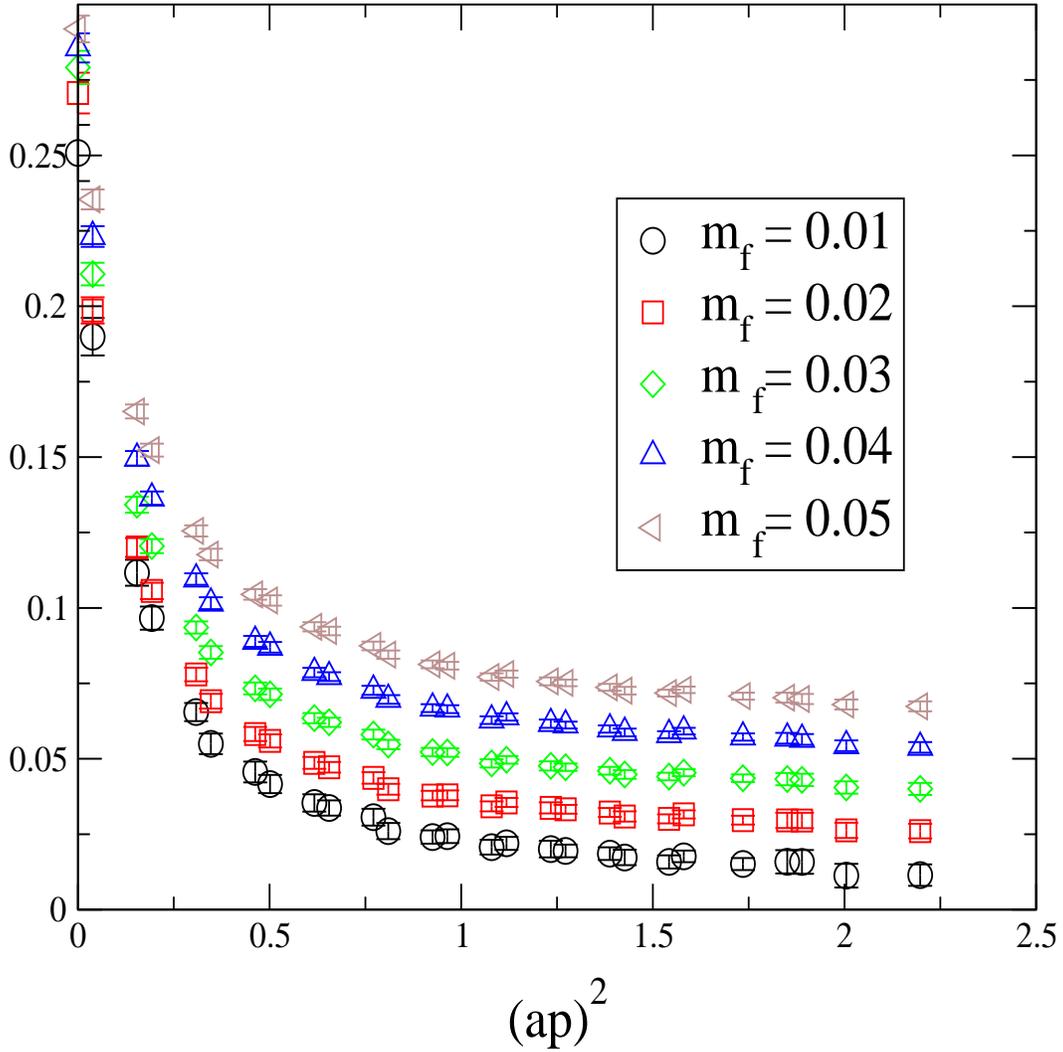}}
\caption{A plot of $\frac{1}{12}{\rm Tr}\left( S^{-1}_{latt} \right)$ versus
$(ap)^2$ showing that for moderate values of $(ap)^2$ the effects of explicit
chiral symmetry breaking are small.}
\label{fig:mass}
\end{center}
\end{figure}

\begin{figure}[!htb]
\begin{center}
\epsfxsize=\hsize
\mbox{\epsfig{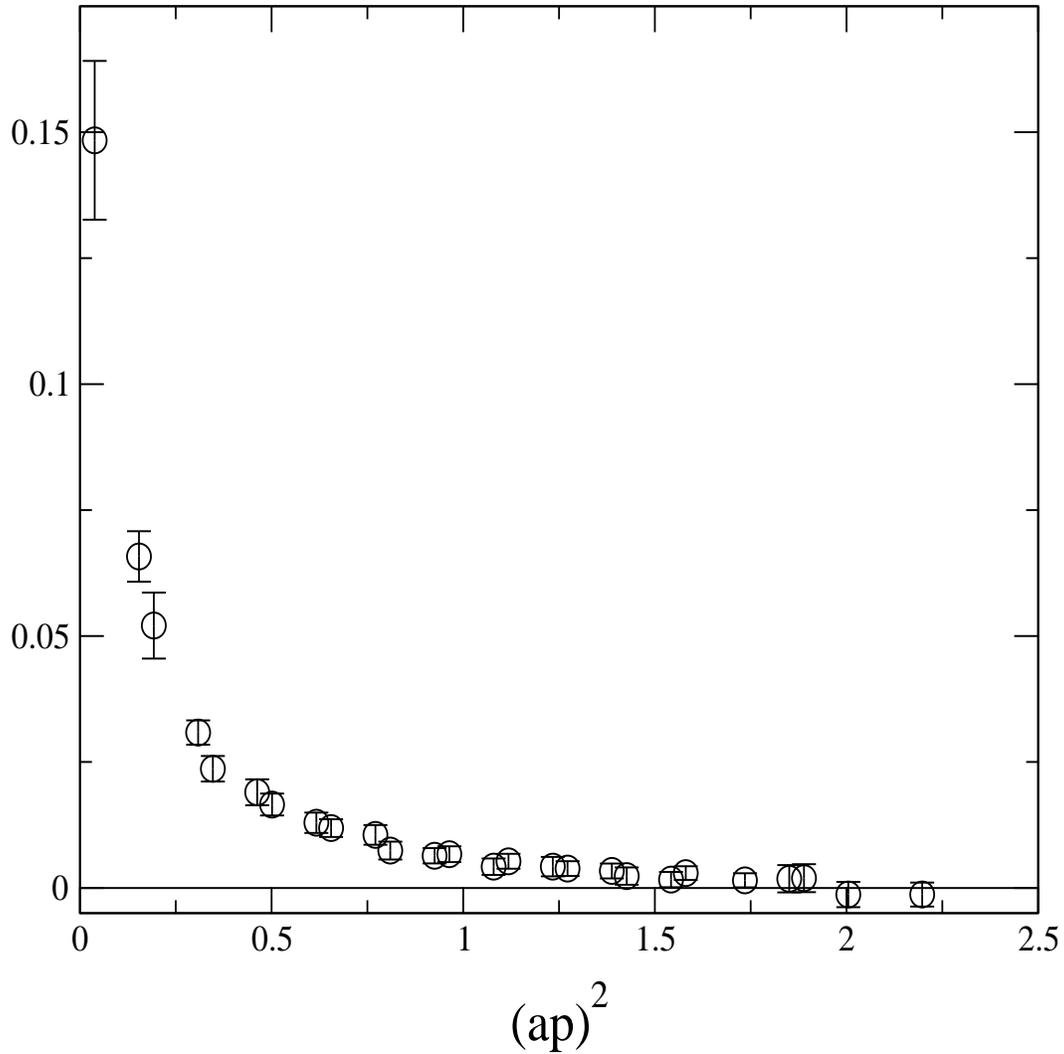}}
\caption{The value  of $\frac{1}{12}{\rm Tr}\left( S^{-1} \right)$
extrapolated to $m_f=0$ vs $(ap)^2$. For moderate $(ap)^2$ the extrapolated
value is zero within errors, showing that the residual mass is small.}
\label{fig:resmas}
\end{center}
\end{figure}

\begin{figure}[!htb]
\begin{center}
\epsfxsize=\hsize
\mbox{\epsfig{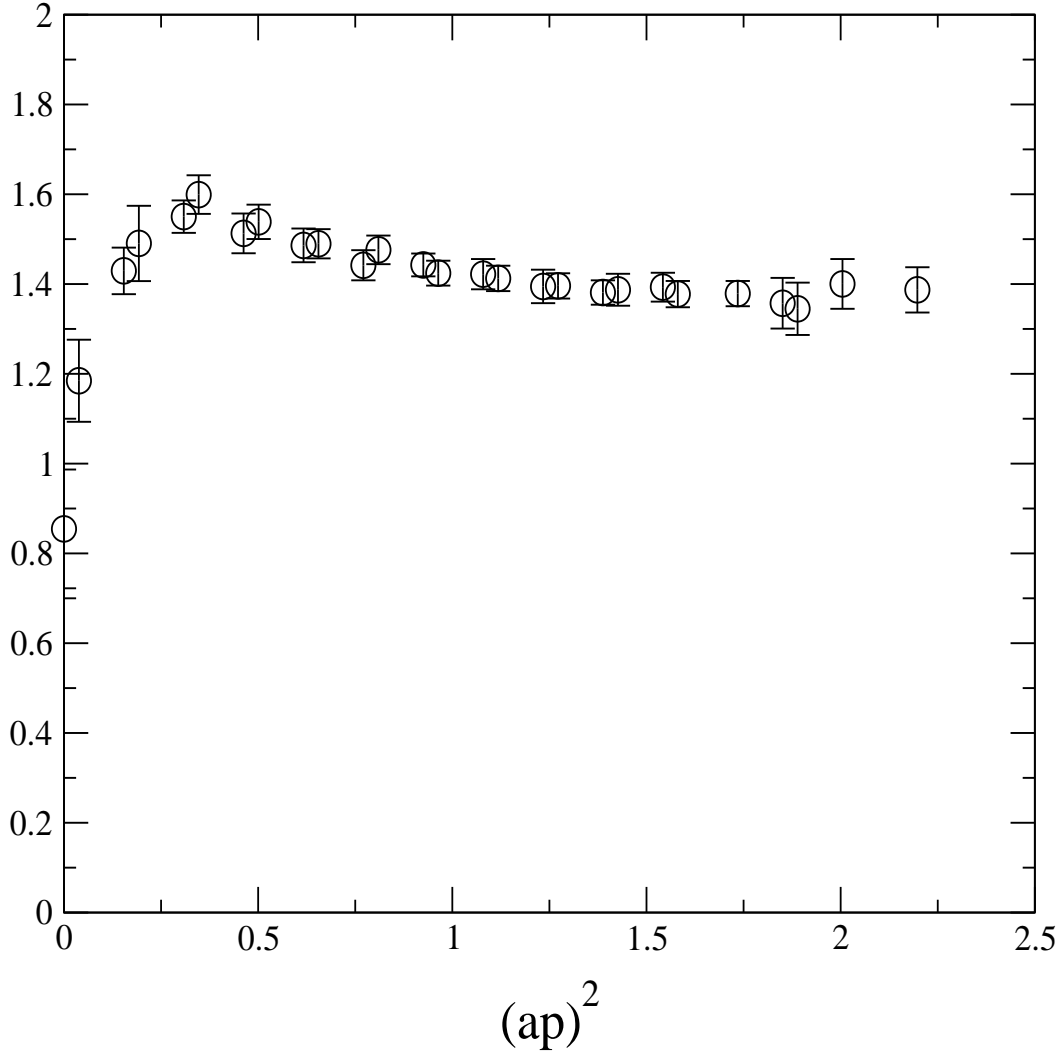}}
\caption{$Z_m Z_q$ calculated from  the slope of $\frac{1}{12}{\rm Tr}\left(
S^{-1} \right)$ versus $m_f$ plotted as a function of $(ap)^2$.}
\label{fig:zmzq}
\end{center}
\end{figure}

In \Fig{fig:mass} we
plot the left-hand side of \Eq{zq:trin} versus $(ap)^2$ for a variety of
values for $m_f$.  As can be seen, for domain wall
fermions this quantity approaches a constant value for moderately large
values of $(ap)^2$. Also, it is encouraging that while
at low momenta the effects of spontaneous chiral symmetry are visible, there
is no evidence to suggest appreciable effects from explicit chiral symmetry
breaking. In particular, there is no evidence of a large additive mass
renormalisation. This is visible in \Fig{fig:resmas}, which shows the
result of a linear extrapolation of the data to the point $m_f=0$.
\Fig{fig:zmzq} shows the slope of this extrapolation, which from \Eq{zq:trin}
is $Z_mZ_q$ at large $(ap)^2$.

\subsection{Extracting $Z_q$ from the Propagator}
\label{sect:zqpr}
The extraction of $Z_q$ from the propagator via \Eq{zq:ward} is numerically
challenging due to the need for a discrete derivative to be calculated. A much
simpler method \cite{Martinelli:1995ty}, is to calculate
\be
{Z'}_q = -i \frac{1}{12}\sum_\mu \frac{ap_\mu}{(ap)^2} 
\mathrm{Tr} \left( \gamma_\mu S^{-1}_{\mathrm{latt}}(ap) \right) \, .
\label{num:zqprime}
\ee  
This quantity may then be related to $Z_q$ by a perturbative
matching calculation performed in the continuum \cite{Franco:1998bm}.

On the other hand, the use of ${Z'}_q$ to determine $Z_q$ introduces
significant $O(a^2)$ errors through the choice of how the discretised momenta are defined. 
If we were to
replace in \Eq{num:zqprime} the definition of the lattice momentum
in \Eq{num:mom} by, for example,
\be  
\bar{p}_\mu = \frac{1}{a}\sin p_\mu a \, , 
\ee
then the resulting ${Z'}_q$ would differ from that given in \Eq{num:zqprime} by 
$O(a^2)$
because the trace includes an explicit factor of $p_\mu \gamma_\mu$.  One can
estimate the size of this error by using various definitions for the lattice
momentum in the analysis.  As will be shown later in
Sec.\ \ref{sect:zq_prop}, this uncertainty is roughly
10--20\%.  In Sections VIII and IX we introduce two methods for computing
$Z_q$ which avoid this uncertainty.

\section{Flavour Non-Singlet Fermion Bilinears}
\label{sect:bil}
\subsection{Introduction}
In the following the renormalisation of the flavour non-singlet
fermion bilinear operators will be considered. To simplify notation
explicit quark flavours ($u$ and $d$) will be used in the following 
equations. The most general fermion bilinear may be written as
$\overline{u} \Gamma_i d$ with
\be
\Gamma_i \in \left\{ 1 , \gamma_\mu , \gamma_5, \gamma_\mu
\gamma_5, \sigma_{\mu \nu} \right\} \, ,
\ee
where $i$ represents whatever indices the gamma matrices have.
The renormalised operator is defined as
\be
\left[ \overline{u} \Gamma d \right]_{{\rm ren}} 
= Z_{\Gamma} \left[ \overline{u} \Gamma d \right]_{0}
\, .
\ee
The factor $Z_{\Gamma}$ is  fixed in the RI scheme by
defining the unrenormalised, amputated vertex function
\be
\Pi_{\Gamma,0}(p,q)=
\frac{1}{V}
\int d^4z \, d^4x_1 \, d^4x_2 \,  
e^{-ip.x_1+iq.x_2} \,
\langle \, \left[ \overline{u} \Gamma d \right]_0 (z) \, u_0(x_1) \, \overline{d_0}(x_2) \, 
\rangle_{{\rm AMP}} ,
\ee
the corresponding renormalised, amputated vertex function
\be
\Pi_{\Gamma,{\rm ren}}(p,q)= {Z_\Gamma \over Z_q} \Pi_{\Gamma,0}(p,q)
\ee
and requiring that
\be
\Lambda_{\Gamma_i, \mathrm{ren}} (p,p)_{p^2=\mu^2} =
\frac{1}{ \mathrm{Tr} \left( \sum_i \Gamma_i \Gamma_i \right)} \mathrm{Tr}
\left( \, \sum_i \Gamma_i
\Pi_{\Gamma_i, \mathrm{ren}} (p,p)
\right)_{p^2 = \mu^2}
= 1 \, .
\label{bil:ri}
\ee
Note, the corresponding, unrenormalised vertex amplitude is defined by
\be
\Lambda_{\Gamma,0} = \frac{1}{ \mathrm{Tr} \left( \sum_i \Gamma_i \Gamma_i \right)} \mathrm{Tr}
\left( \, \sum_i \Gamma_i
\Pi_{\Gamma_i,0} (p,p)
\right) \, ,
\ee
so that 
\be
\Lambda_{\Gamma_i, \mathrm{ren}} (p,p) = \frac{Z_{\Gamma}}{Z_q}
\Lambda_{\Gamma_i, 0} (p,p) \, .
\label{zcond}
\ee 
While \Eq{zcond} completely defines a procedure for
calculating a renormalisation factor for the bilinear operator
of interest, practically we need to be able to use a renormalisation
condition that allows us to match to perturbative calculations. In general
the value of $\Lambda_{\Gamma_i, 0} (p,p)$ has contributions from
intrinsically non-perturbative effects (such as those due to propagating
pions) that perturbative calculations do not include. As we are interested in
the value of the renormalisation factors in the perturbative regime we  
either apply the renormalisation condition at a high enough momenta
such that the non-perturbative effects are suppressed, or remove such
effects from the data and in the following that is what we will do. We will
reserve the notation $Z_i$, $i \in\{ q,S,P,T,A,V\}$ 
for the renormalisation factors in the perturbative regime.

\subsection{$Z_A$ and $Z_V$}

A (partially) conserved current that is normalised in a fashion which is
consistent with the usual Ward-Takahashi identities will undergo no
renormalisation and the corresponding $Z_\Gamma$ will be unity.  In
particular, for domain wall fermions and the RI renormalisation scheme
specified by Eqs.~\ref{zq:ward} and \ref{bil:ri} and imposed at high-momentum,
$\mu \gg \Lambda_{QCD}$, we expect 
\be 
Z_{\cal A} = Z_{\cal V} = 1.  
\ee
However, on the lattice the (partially) conserved currents are not local and
it is frequently more convenient to work with their local
counterparts. Provided that these are related by a chiral transformation one
still has \be Z_A = Z_V \, .  \ee This does not, however, mean that
$\Lambda_A$ must equal $\Lambda_V$, and several mechanisms exist for splitting
them away from each other at low energies.

Even if there are no significant effects from explicit chiral symmetry
breaking, the effects of spontaneous chiral symmetry breaking must be taken
into account.  Consideration of the operator product expansion, to 
lowest order in powers of $1/p^2$, shows that $\Lambda_A$ and
$\Lambda_V$ may get contributions from terms of the form
\be
\frac{m_{\mathrm{ren}}^2}{p^2}
\label{bil:mqq}
\ee 
and
\be
\frac{m_{\mathrm{ren}} \langle \overline{q}q \rangle}{p^4} \, .
\label{bil:mqq2}
\ee 
Since such terms, by their very nature, stem from chiral symmetry
breaking they are not constrained to enter $\Lambda_A$ and 
$\Lambda_V$ with the same weight. 
At large momenta these terms are suppressed and do not effect the extraction of
$Z_A$ and $Z_V$.

If the action being used explicitly breaks chiral symmetry $Z_A$ and $Z_V$
need not be equal, but their ratio will still be scale independent. This means
that while $\Lambda_A$ and $\Lambda_V$ need not approach one another at high
momenta, their ratio should become scale independent for large
enough momenta.

While we want to work in the chiral limit for the extraction of the
renormalisation factors it is also worthwhile to consider
what the mass dependence of $\Lambda_A$ and $\Lambda_V$ should be (especially
as we wish to extract the chiral limit from data measured in finite mass
simulations). Requiring that the ``generalised'' symmetry
introduced in \Eq{dwf:gen} is satisfied,  constrains the mass dependence
to either be of the same form as \Eq{bil:mqq} (a single power of mass
multiplying something that breaks chiral symmetry - and therefore by the
argument of the previous paragraph damped with momentum) or 
proportional to a second or higher power of mass. In the latter case any effect
on our data should be negligible. 

\begin{figure}[!htb]
\begin{center}
\epsfxsize=\hsize
\mbox{\epsfig{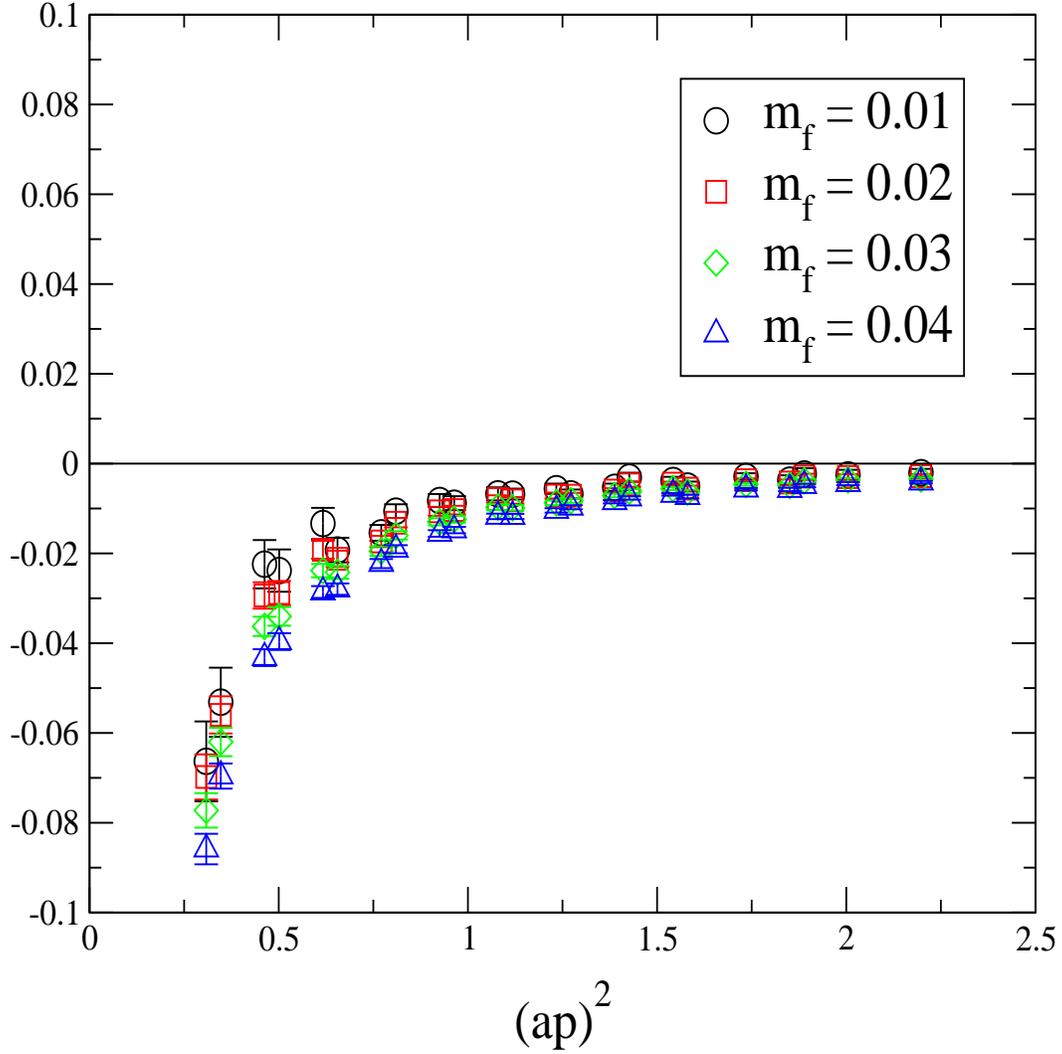}}
\caption{A plot of $\Lambda_A - \Lambda_V$ versus $(ap)^2$, showing
that there is no significant difference between $Z_A$ and $Z_V$, 
even for moderate values of $(ap)^2$.}
\label{fig:amv}
\end{center}
\end{figure}

\begin{figure}[!htb]
\begin{center}
\epsfxsize=\hsize
\mbox{\epsfig{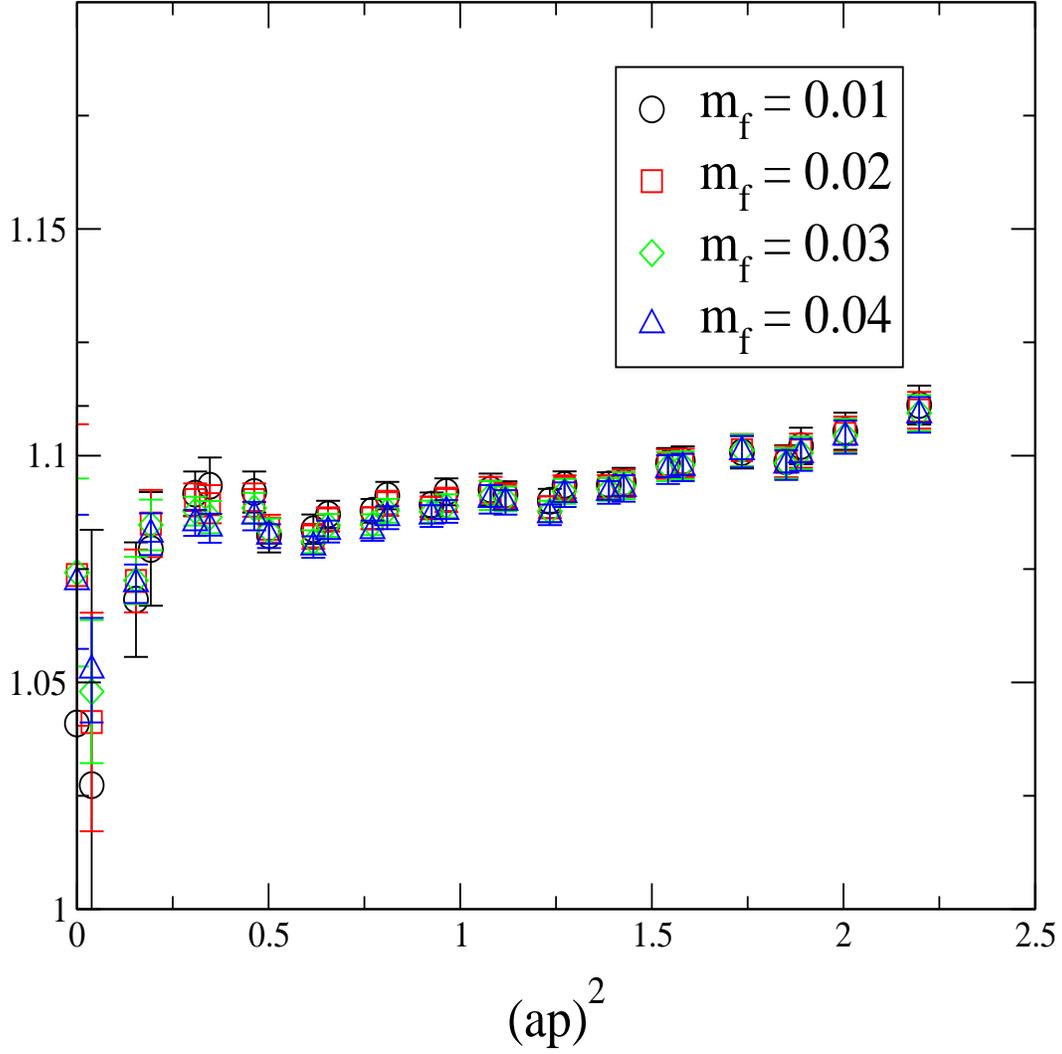}}
\end{center}
\caption{ A graph of $\frac{1}{2}\left\{\Lambda_A + \Lambda_V\right\}$ versus
$(ap)^2$, which up to lattice artifacts, gives $Z_A/Z_q$ and $Z_V/Z_q$.}
\label{fig:apv}
\end{figure}

The above considerations suggest looking at the quantity
\be
\Lambda_A - \Lambda_V \, .
\ee
This is shown in \Fig{fig:amv} and, as with the case for the quark propagator,
while the effects of non-perturbative breaking terms are visible at low
momenta, they are damped at higher momenta. There is no significant signal for
effects from explicit chiral symmetry breaking, since $\Lambda_A- \Lambda_V$
is tending to zero. At momenta of interest, there also seems to be
no significant splitting due to non-perturbative effects with the
difference between $\Lambda_A$ and $\Lambda_V$ being less than
$1\%$ in the chiral limit at $(ap)^2=0.8$ and smaller for momenta above
this. This being so it is sensible to use the quantity
$\frac{1}{2}\left(\Lambda_A + \Lambda_V\right)$ for the extraction of
both $Z_A/Z_q$ and $Z_V/Z_q$ to increase the statistical accuracy. 
This is shown in \Fig{fig:apv}.
%

\subsection{ $Z_S$ and $Z_P$ }
\begin{figure}[!htb]
\begin{center}
\epsfxsize=\hsize
\mbox{\epsfig{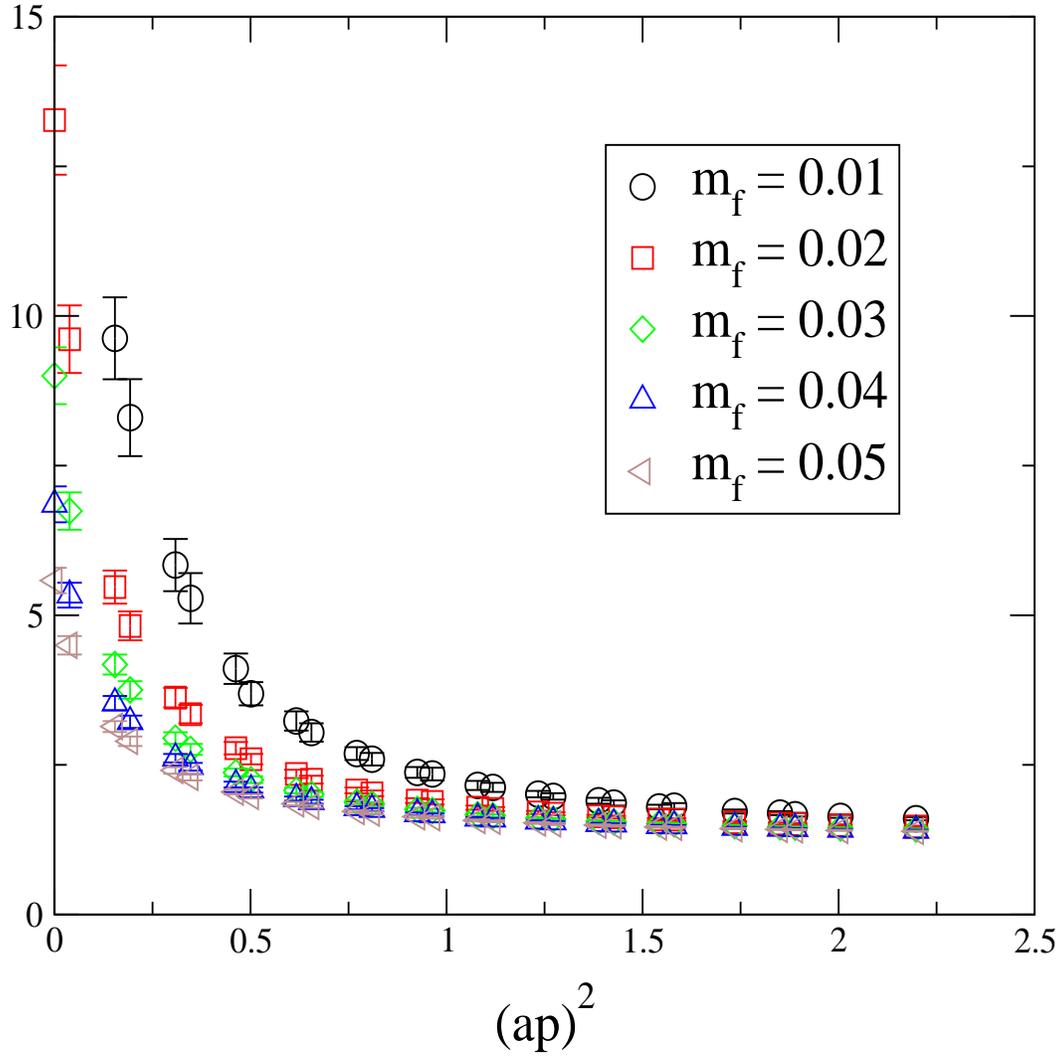}}
\caption{$\Lambda_P$ versus $(ap)^2$ for several values of $m_f$, showing
that the $1/p^2$ pole is more pronounced for small $m_f$.}
\label{fig:zppole}
\end{center}
\end{figure}

For a theory with chiral symmetry the
RI scheme preserves the well known $\overline{MS}$ relations
\bea
Z_S &=& Z_P \\ 
Z_m &=& \frac{1}{Z_S} \, .
\eea
If the potential for explicit chiral symmetry breaking is taken
into account, these equalities cease to be valid, but the
quantities $Z_S/Z_P$, $Z_SZ_m$ and $Z_m Z_P$ are expected
to be scale independent.
However, lattice studies using both the Wilson and staggered actions
have shown that the ratio $\Lambda_P/\Lambda_S$,
which in perturbation theory would be equal to $Z_S/Z_P$
and therefore might be expected to be momentum independent up to
small corrections, is strongly momentum and mass dependent,
with the bulk of this dependence arising from $\Lambda_P$.
It is instructive to consider the source of this discrepancy
\cite{Martinelli:1995ty,Cudell:1998ic}.
We start from the continuum axial Ward-Takahashi identity which is
derived by taking the axial variation of the quark propagator.
This reads
\be
(m_u+m_d)
\int dz
\langle u(x_1) \left[ \overline{u}\gamma_5d \right](z) \overline{d}(x_2) \rangle 
=
\gamma_5 \langle d(x_1) \overline{d}(x_2) \rangle
+
\langle u(x_1) \overline{u}(x_2) \rangle \gamma_5 \, .
\ee
Moving to momentum space gives
\be
(m_u+m_d)
\langle u \left[ \overline{u}\gamma_5d \right] \overline{d} \rangle (p,p) 
=
\gamma_5 \langle d \overline{d} \rangle (p) 
+
\langle u\overline{u} \rangle (p) \gamma_5\, ,
\ee
which in the $m_u \rightarrow m_d \equiv m$ limit gives
\be
2 m \langle u \left[\overline{u}\gamma_5 d\right] \overline{d} \rangle (p,p) 
=
\left\{ \gamma_5,S(p) \right\} \, .
\ee
Amputating and tracing with $\gamma_5$ yields
\be
m \Lambda_P (p,p) 
=
\frac{1}{12}\mathrm{Tr} \left( S^{-1}(p) \right) \, .
\ee
Neglecting all lattice artifact terms except the additive mass renormalisation
(which is justified by
the discussions in Section~\ref{sect:renprop}), this leads
to an approximate expression for $\Lambda_{P,{\rm latt}}$, including
the first order contribution of spontaneous chiral
symmetry breaking: 
\be
\left( m_f + m_{{\rm res}} \right) \Lambda_{P,{\rm latt}} \left( ap,ap\right)
=
\cdots +
\frac{a^2\langle \overline{q} q \rangle}{(ap)^2} C_1 Z_q
+\, Z_m Z_q \left( m_f + m_{res} \right) \, ,
\label{bil:zpp}
\ee
neglecting terms of \order{$(ap)^2$}.
While in the absence of the condensate term this equation 
reduces to $Z_P = 1/Z_m$,
the condensate term, which is clearly visible in \Fig{fig:resmas}, 
gives rise to a pole term of the form 
\be
\frac{\langle \overline{q} q \rangle}
{ \left(m_f +m_{res}\right)(ap)^2} C_1 Z_q
\ee 
in $\Lambda_{P,{\rm latt}}$.
\Fig{fig:zppole} shows the data for $\Lambda_{P,{\rm latt}}$
with this effect clearly visible, with the rise for small $(ap)^2$ becoming
much more pronounced as the mass decreases. 

\begin{figure}[!htb]
\begin{center}
\epsfxsize=\hsize
\mbox{\epsfig{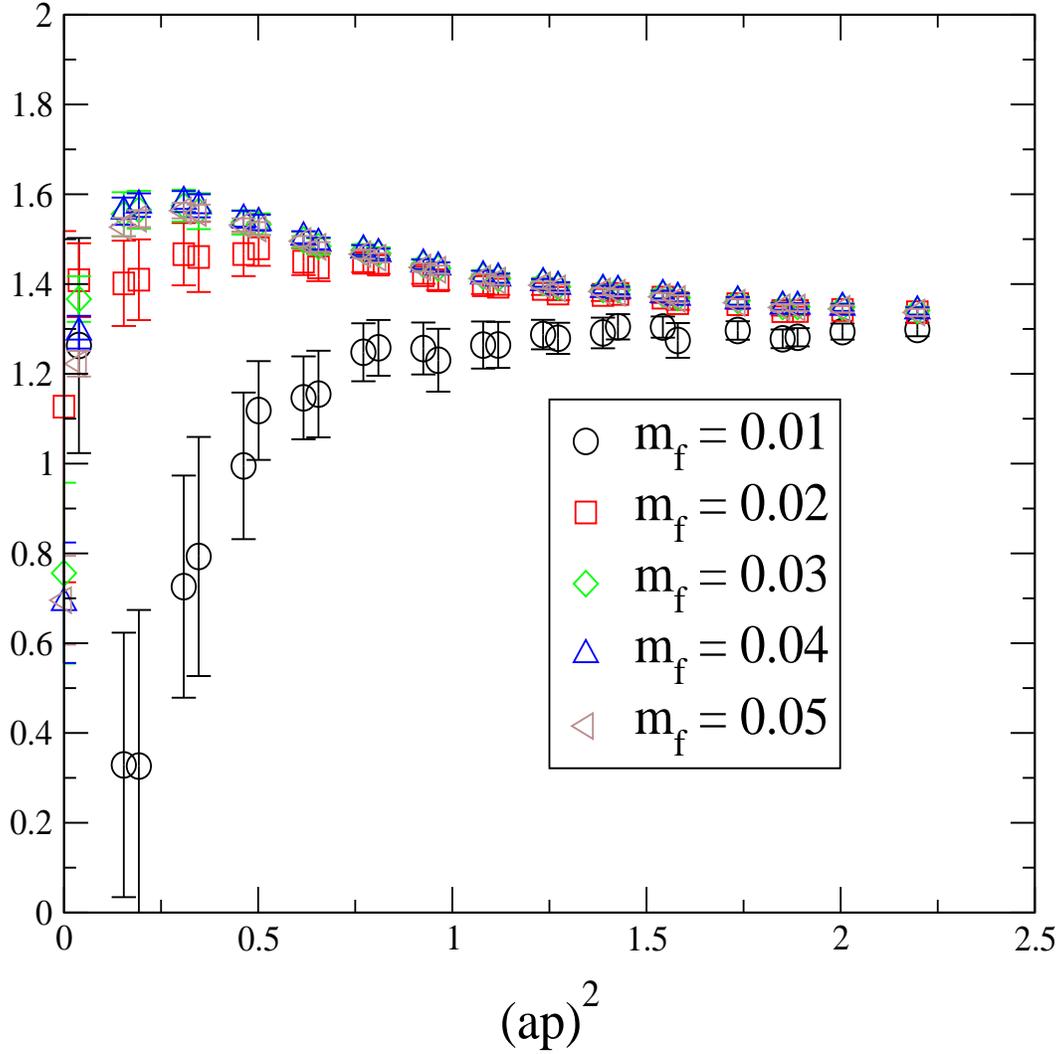}}
\caption{A plot of $\Lambda_{S,\mathrm{latt}}$ versus $(ap)^2$ for
several masses.
The mass pole can be clearly seen for small momenta and is attributable to 
zero-mode effects. }
\label{fig:zspole}
\end{center}
\end{figure}

A similar argument may be put forward for $\Lambda_S$. In this case it
is the Ward-Takahashi identity arising from a vector rotation of the
fields.  
\be
(m_u- m_d)
\int dz
\langle u(x_1) \left[ \overline{u} d \right](z) \overline{d}(x_2) \rangle 
=
\langle d(x_1) \overline{d}(x_2) \rangle
-
\langle u(x_1) \overline{u}(x_2) \rangle
\ee
Moving to momentum space, this gives
\be
(m_u- m_d)
\langle u \left[ \overline{u}d \right] \overline{d} \rangle (p,p) 
=
\langle d \overline{d} \rangle (p) 
-
\langle u\overline{u} \rangle (p) \, ,
\ee
which in the $m_u \rightarrow m_d$ limit tends to
\be
\langle u \left[\overline{u}d\right] \overline{d} \rangle (p,p) 
=
-\frac{\partial}{\partial m} S(p) \, .
\ee
Finally, amputating and tracing gives 
\be
\Lambda_S
=
\frac{1}{12}\frac{\partial \, {\rm Tr} \left[ S(p)^{-1} \right]}{\partial m} \, .
\ee
Note that this relation should be exact for domain wall fermions (with $m=
m_f$) for any value of $L_s$. Using Eq~\ref{zq:trin}, and noting that
both the residual mass and the renormalisation factors should be independent
of $m_f$, gives the approximate expression
\be
\Lambda_{S,{\rm latt}} =  Z_m Z_q  + \frac{C_1Z_q}{(ap)^2} 
\frac{\partial a^3 \langle \overline{q} q \rangle}{\partial (am_f)} \, .
\label{bil:zsp}
\ee
If the mass dependence of $\langle\overline{q}q \rangle$, for small
masses, is proportional to only positive powers of the mass, then the
second term in \Eq{bil:zsp} is almost certainly unimportant as it
is suppressed in exactly the region of parameter space in which we are
working: large momenta and small masses. (The effect might
be larger than naively expected, as
$\langle\overline{q}q\rangle$ is quadratically divergent in the
lattice spacing.)  It is necessary, however, to consider the effects of
fermionic zero-modes on $\langle\overline{q}q\rangle$.  Assuming a
theory with chiral symmetry, the spectral decomposition of the
quark propagator leads to an expected form for
$\langle\overline{q}q\rangle$, on a single configuration C, of
\bea
-\langle \overline{q} q \rangle_{C}
&=&
Tr \left[ \left( \slash{D} + m \right)^{-1} \right]
\\ 
&=&
\frac{n_0}{mV}
+  
\frac{1}{V}\sum_{\lambda_n > 0 }
\frac{1}{m + i\lambda_n} \, ;\, n_0 \ge 0 .
\label{bil:pbp}
\eea
where $n_0$ is the number of fermionic zero-modes, $V$ is the 
four dimensional space-time volume
and the $\lambda_n$ are such that $\slash{D} \psi_n = i\lambda_n \psi_n$.
 The number of such zero
modes should grow more slowly than the volume, and so the first term in
\Eq{bil:pbp} will vanish in the infinite volume limit.  However, for the
lattice parameters used for this simulation the effects of zero-modes have
been found to be noticeable in both $\langle\overline{q}q\rangle$ and 
hadronic spectrum
calculations~\cite{Blum:2000kn} and so must be considered for the present case. 
Comparing \Eq{bil:zsp} and \Eq{bil:pbp} shows that, 
as $m\rightarrow 0$ for fixed momentum, $\Lambda_S$ gets a 
large contribution from zero-modes of the form
\be
-\frac{1}{m^2} \langle n_0 \rangle \frac{C_1 Z_q}{p^2} \, ,
\label{bil:zspole}
\ee
that must be subtracted before $Z_S/Z_q$ may be calculated from 
\Eq{bil:ri}. \Fig{fig:zspole} shows our data for $\Lambda_{S,\mathrm{latt}}$. 
While the effect of the condensate term is smaller than that for $Z_P$, 
it is noticeable for the lighter masses.

\subsection{Fitting the Pole Terms}

Considering \Eq{bil:zspole} and moving to lattice notation, the method for
extracting $Z_S$ from $\Lambda_S$ becomes clear. Working at a fixed momentum,
$\Lambda_S$ may be fitted to the form
\be
\Lambda_{S,{\rm latt}}
 = 
\frac{c_{1,S}}{\left(am_f+ {\cal O }(am_{{\rm res}})\right)^2} + c_{2,S} +
c_{3,S} (am_f)^2 \, ,
\ee
with $Z_q/Z_S$ being given by $c_{2,S}$. While one might naively expect the
denominator
in the above equation to be $am_f + am_{\mathrm{res}}$, as shown in
Ref.~\cite{Blum:2000kn}
the residual chiral symmetry breaking effects that appear in the pole
term in $\overline{q}q$ are not parameterised precisely by 
$am_{\mathrm{res}}$ since the singular behavior of the pole enhances
what are expected to order $a^2$ variations in the quantity $am_{\mathrm{res}}$. 
One only knows that the residual chiral symmetry breaking effects are
of \order{$am_{\mathrm{res}}$}. However, for the pole subtractions in this
paper we have used precisely $am_f + am_{\mathrm{res}}$. This is justified
since the statistical errors on our data are such that the fits are
insensitive to the exact value of $m_{\mathrm{res}}$.

\begin{figure}[!htb]
\begin{center}
\epsfxsize=\hsize
\mbox{\epsfig{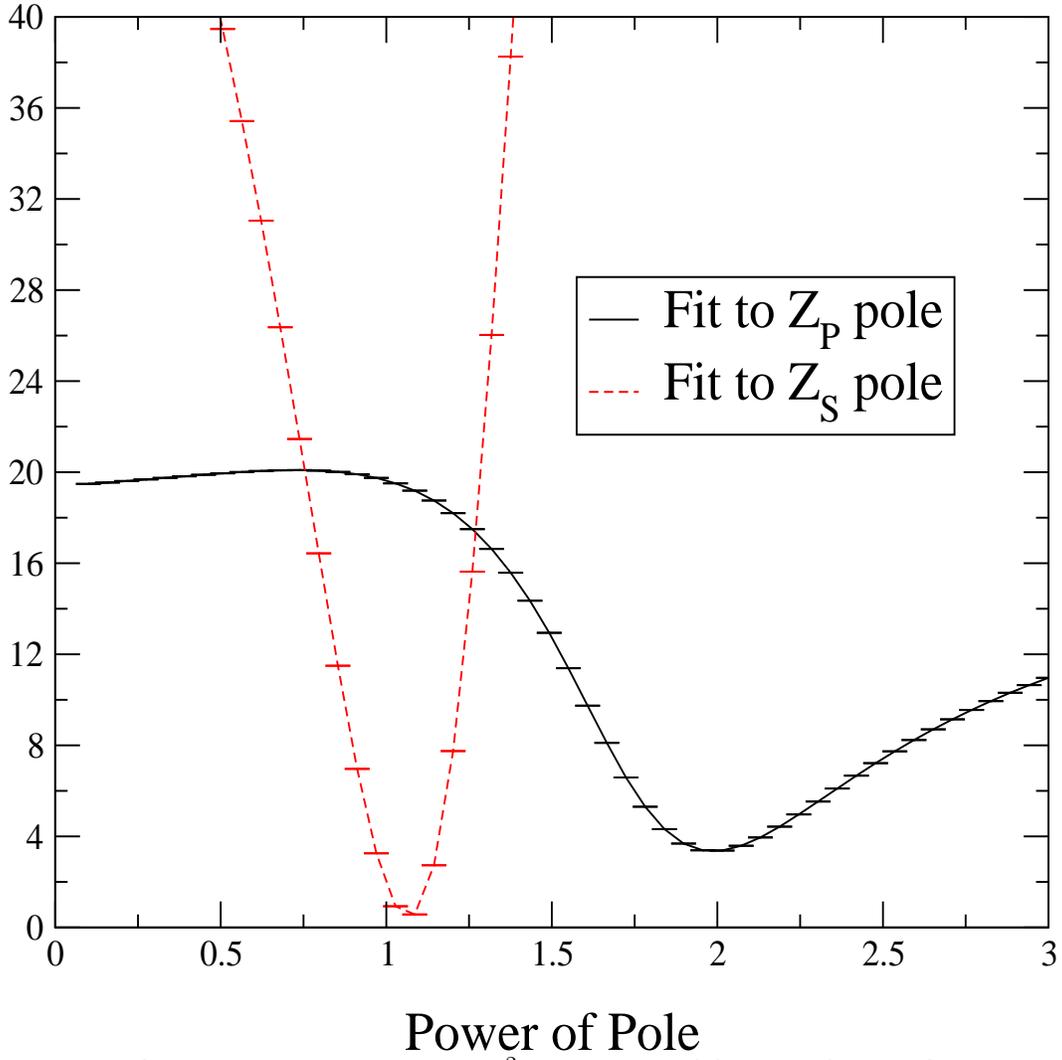}}
\caption{This figure displays the average $\chi^2$ per degree of freedom for
the fits used to determine the power of the mass poles in
$\Lambda_S$ and $\Lambda_P$, clearly showing their double and single
poles, respectively.}
\label{powerfit}
\end{center}
\end{figure}

\begin{figure}[!htb]
\begin{center}
\epsfxsize=\hsize
\mbox{\epsfig{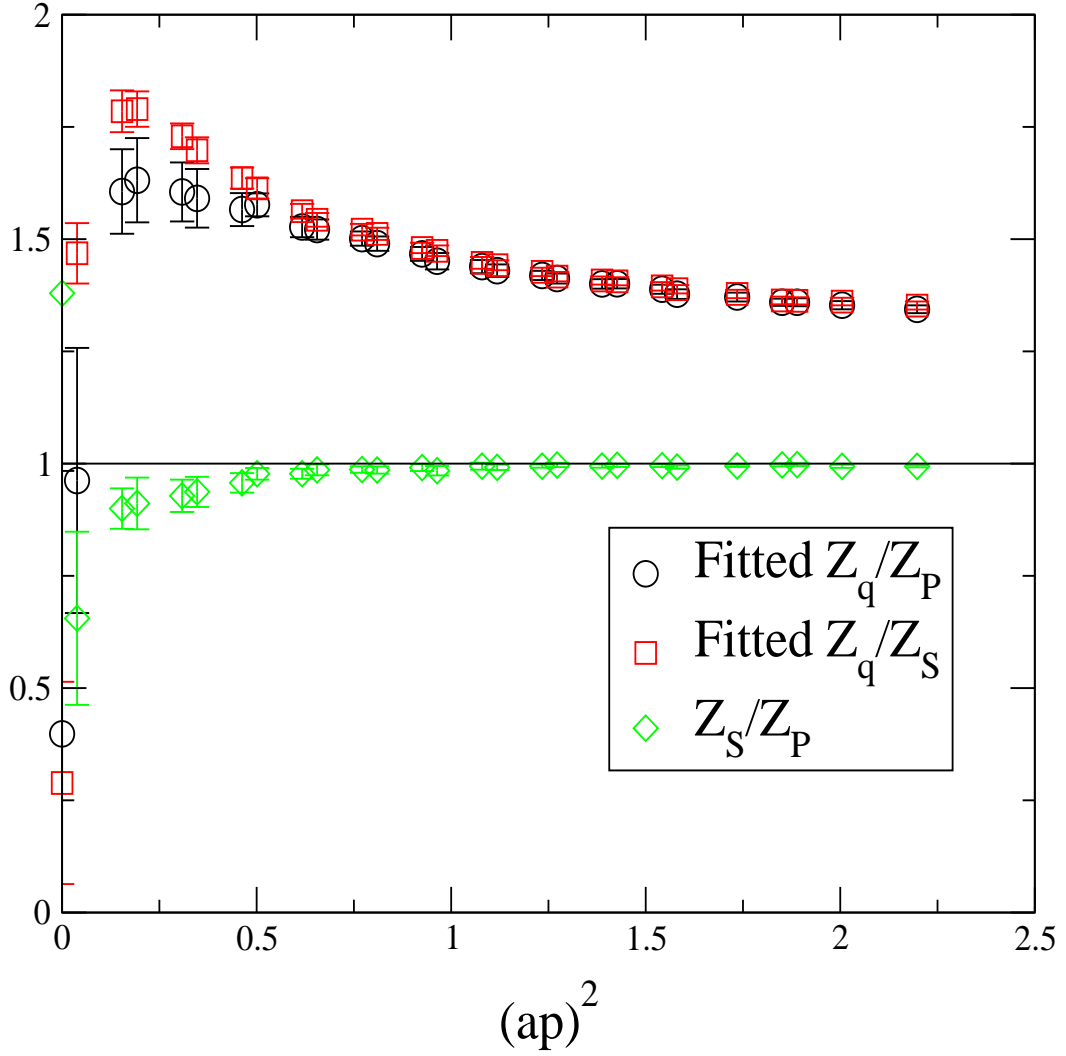}}
\caption{A comparison of $Z_S/Z_q$ and $Z_P/Z_q$ as extracted from $\Lambda_S$
and $\Lambda_P$ after pole subtraction.}
\label{zszpcomp}
\end{center}
\end{figure}

\begin{figure}[!htb]
\begin{center}
\epsfxsize=\hsize
\mbox{\epsfig{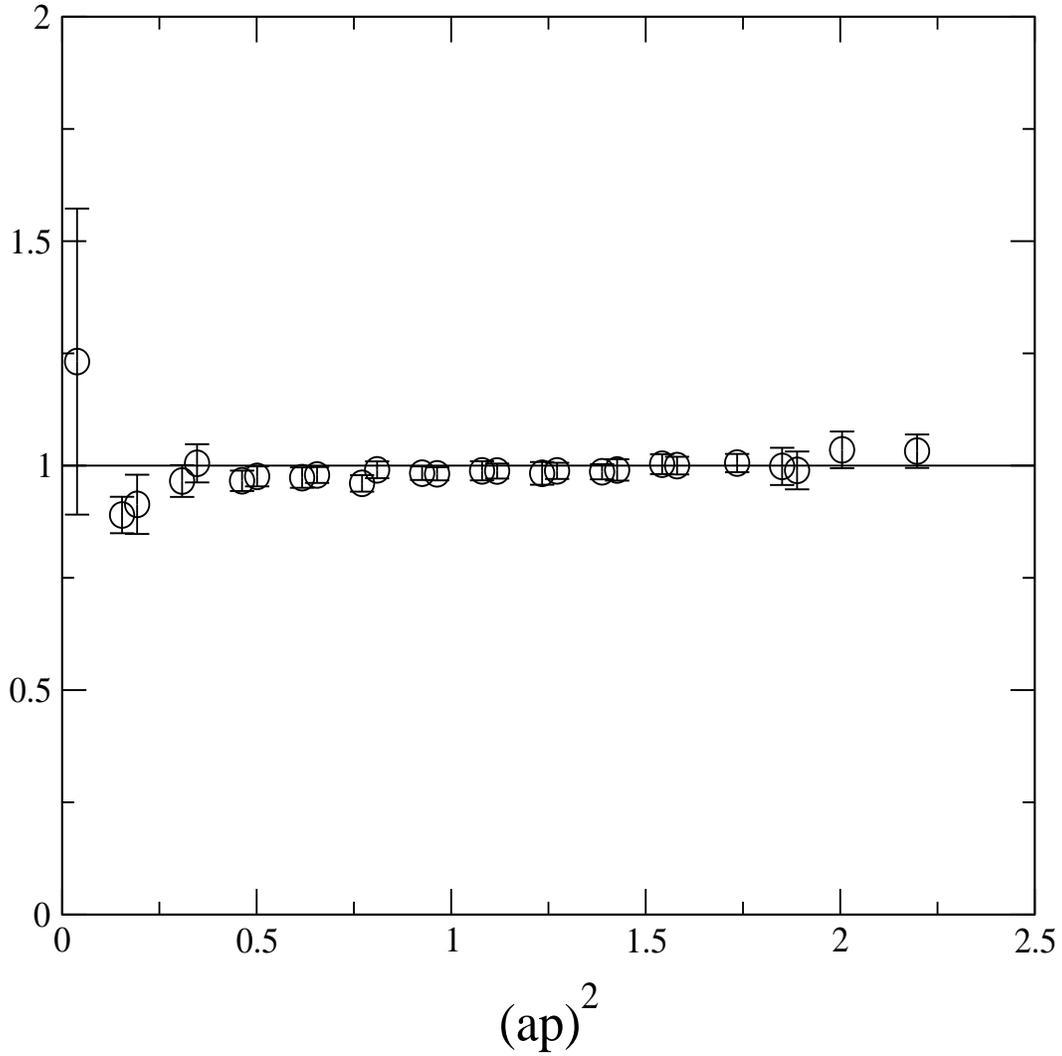}}
\caption{The product $Z_PZ_m$ calculated by combining $Z_P/Z_q$ from pole
subtraction with the trace of the inverse propagator. This product is clearly 
unity within errors.}
\label{fig:zpzm}
\end{center}
\end{figure}

The situation for
the $Z_P$ extraction is slightly more complicated.
Examining \Eq{bil:zpp} and \Eq{bil:pbp}, shows that $\Lambda_{P}$
should have a double pole of the form
\be
\Lambda_{P,{\rm latt}}
 = 
\frac{c_{1,P}}{(am_f+am_{{\rm res}})^2} 
+ 
\frac{c_{2,P}}{(am_f+am_{{\rm res}})} 
+
c_{3,P} +
c_{4,P} ( am_f )^2 \, ,
\label{eq79}
\ee 
with $c_{3,P}$ being equal to $Z_q/Z_P$ and the quadratic mass pole due to
zero-mode effects in $\langle\overline{q}q\rangle$. For simplicity,  we have 
again used $am_f + am_{\mathrm{res}}$ in the first term of the right-hand 
side in \Eq{eq79}. For practical purposes, however, the need to fit to the
quadratic term may be avoided by working with $m_f \ge 0.02$. 
Good evidence that the above fitting forms are correct is shown in
\Fig{powerfit}. This shows the average, over all the momenta in the
range $0.5<\left( ap\right)^2< 2.0$, of the $\chi^2$ per degree of freedom for
a correlated fit to the above forms for $m_f$ dependence with the power of the
pole treated as a free parameter.  One sees that a single pole is favored for
the $Z_P$ case while a double is identified in the fit for $Z_S$.  
Further evidence is provided by considering
the resulting values for $Z_S/Z_q$ and $Z_P/Z_q$. \Fig{zszpcomp} shows a
comparison between the extracted values of these two quantities. As chiral
symmetry would predict for $Z_S/Z_q$ and $Z_P/Z_q$, the two quantities
coincide at large momenta. This provides an excellent test of both the
extraction method and the chiral properties of domain wall fermions. This can
be further seen by comparing $Z_P$ and $Z_m$ (as calculated from the trace of
the inverse propagator), which is shown in \Fig{fig:zpzm}. This product is
clearly very close to unity.
\subsection{ $Z_T$ }
\begin{figure}[!htb]
\begin{center}
\epsfxsize=\hsize
\mbox{\epsfig{file=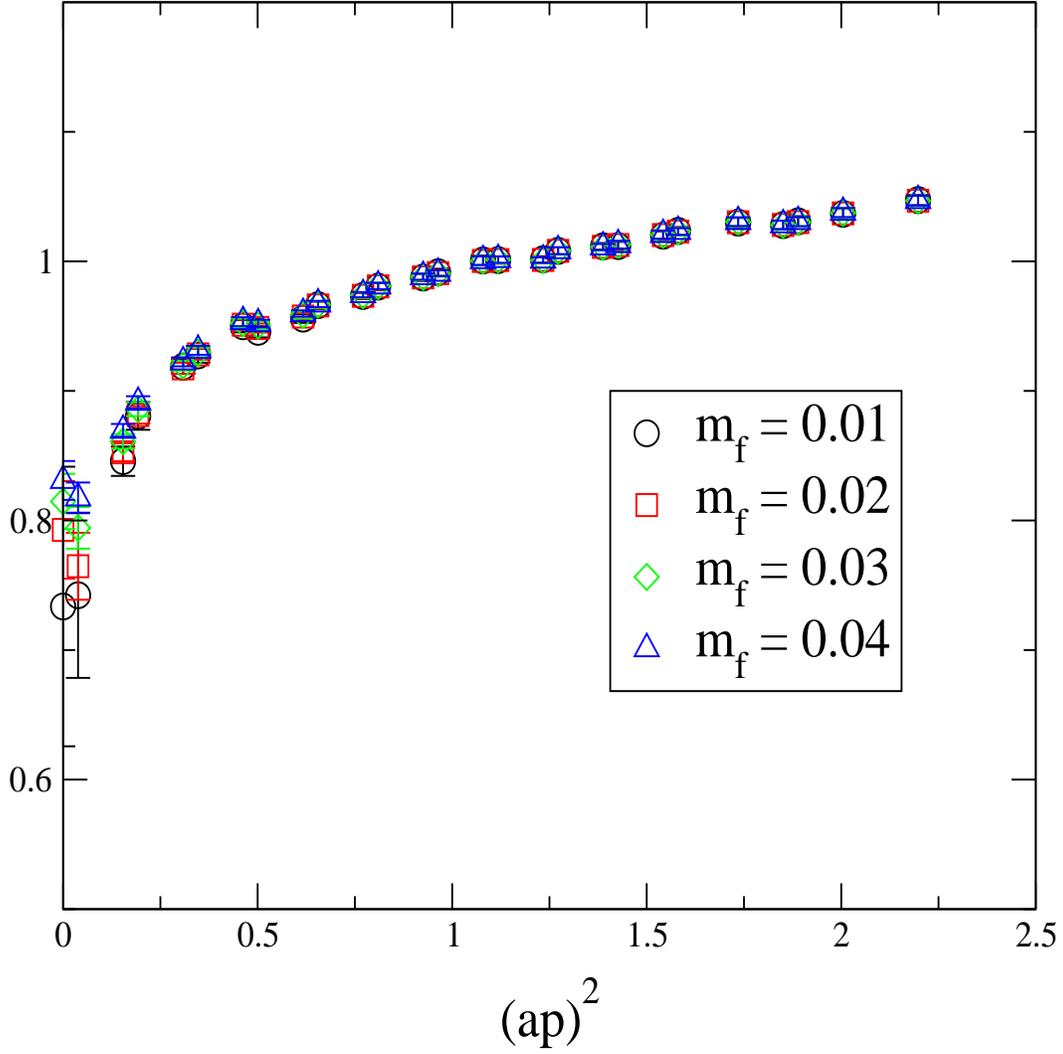,width=14truecm, height=14truecm}}
\caption{This is a graph of $\Lambda_T$ versus $(ap)^2$, from which $Z_T$
will be extracted in Section \ref{sec:rg}. }
\label{fig:zt}
\end{center}
\end{figure}

The tensor density renormalisation, while sometimes neglected 
in bilinear renormalisation coefficient calculations, is a 
quantity of use to current lattice simulations \cite{Blum:2000cf}. 
An extraction of its value will be postponed to Section~\ref{sec:rg},
but for completeness a plot of $\Lambda_T$ is shown in \Fig{fig:zt}.

\section{Extracting $Z_q$ from the Exact Ward Identities.}
\label{sec:zqward}
The vector Ward-Takahashi identity of \Eq{dwf:evec} is exact at finite lattice
spacing.  As such, the renormalisation coefficient for the conserved vector
current defined in \Eq{dwf:cvec}, is equal to unity.  Additionally, the
considerations of Section~\ref{sec:chi_break} show that through first order in
the residual chiral symmetry breaking the extra $J_{5q}^a$ in \Eq{dwf:eveca}
can be completely absorbed into the additive renormalisation of the mass so
that the axial Ward-Takahashi identity, for low-energy physics, takes on the
normal, continuum form. Therefore, the renormalisation factor for the axial
current defined by \Eq{dwf:acur} should also be unity to a good approximation.

The above facts can be used to compute
the quark field renormalisation $Z_q$ 
from \Eq{bil:ri}, as applied the conserved 
vector and axial currents. For the case of the conserved vector
current \Eq{bil:ri} reads:
\be
Z_q^{-1} Z_{{\cal V}} \frac{1}{48} \mathrm{Tr}
(\gamma^\mu \Lambda_{{\cal V}}^\mu(p)) = 1 \,
\ee
which therefore implies
\be
Z_q = \frac{1}{48} {\rm Tr}(\gamma^\mu \Lambda_{{\cal V}}^\mu(p)).
\ee
A similar equation also holds for the conserved axial current.

Because the formulae for the conserved currents contain
fermion fields at separate points on the lattice
as well as summation over all $s$,
the calculation of the matrix elements for these operators can be
very expensive. We used a random source estimator to compute the
part of the sum between $s=1$ to $s=L_s-2$. Also, instead of calculating
all four components of $\Lambda^\mu(p)$ for a given momentum $p$, we
calculate $\Lambda^0(p)$ for momenta related to $p$ by interchange
of its $0$th component with each of the other three. We then
use the equality
\bea
\nonumber
\mathrm{Tr}\left(
\gamma^\mu \Lambda_{\cal V}^\mu(p^0,p^1,p^2,p^3)
\right) 
&=&
\mathrm{Tr}\left(\gamma^0 \Lambda_{\cal V}^0(p^0,p^1,p^2,p^3)\right)\\
\nonumber
&&
+ \mathrm{Tr}\left(\gamma^0 \Lambda_{\cal V}^0(p^1,p^0,p^2,p^3)\right)\\
\nonumber
&&   
+ \mathrm{Tr}\left(\gamma^0 \Lambda_{\cal V}^0(p^2,p^1,p^0,p^3)\right)\\
&&
+ \mathrm{Tr}\left(\gamma^0 \Lambda_{\cal V}^0(p^3,p^1,p^2,p^0)\right)
\label{yuri:rot} \, ,
\eea
to obtain the needed result. As the time to Fourier transform a matrix
element is negligible in comparison with the calculation of the matrix element
itself, this allows us to obtain the result with only a
quarter of running time of direct calculation of $\Lambda_{\cal V}^\mu$.

Because the volume $16^3 \times 32$ used in the simulations is not
symmetric, strictly speaking the above equation is not exact, as the
third component of momentum is not related by symmetry to the first
and second ones.  However, this difference is suppressed by two powers of the
lattice spacing and, in practice, the results obtained for the
last three terms in
\Eq{yuri:rot} all agree within statistical errors for the momenta
used.
We used momenta with the first three integer components no larger than 
$2$ and the last component equal to $0$ or $2$.
The two exceptions were the momenta with integer components
$(2,2,2,0)$ and $(2,2,2,2)$ that were excluded since they would
require usage of momenta $(0,2,2,4)$ and $(1,2,2,4)$.

\begin{figure}[!htb]
\begin{center}
\epsfxsize=\hsize
\mbox{\epsfig{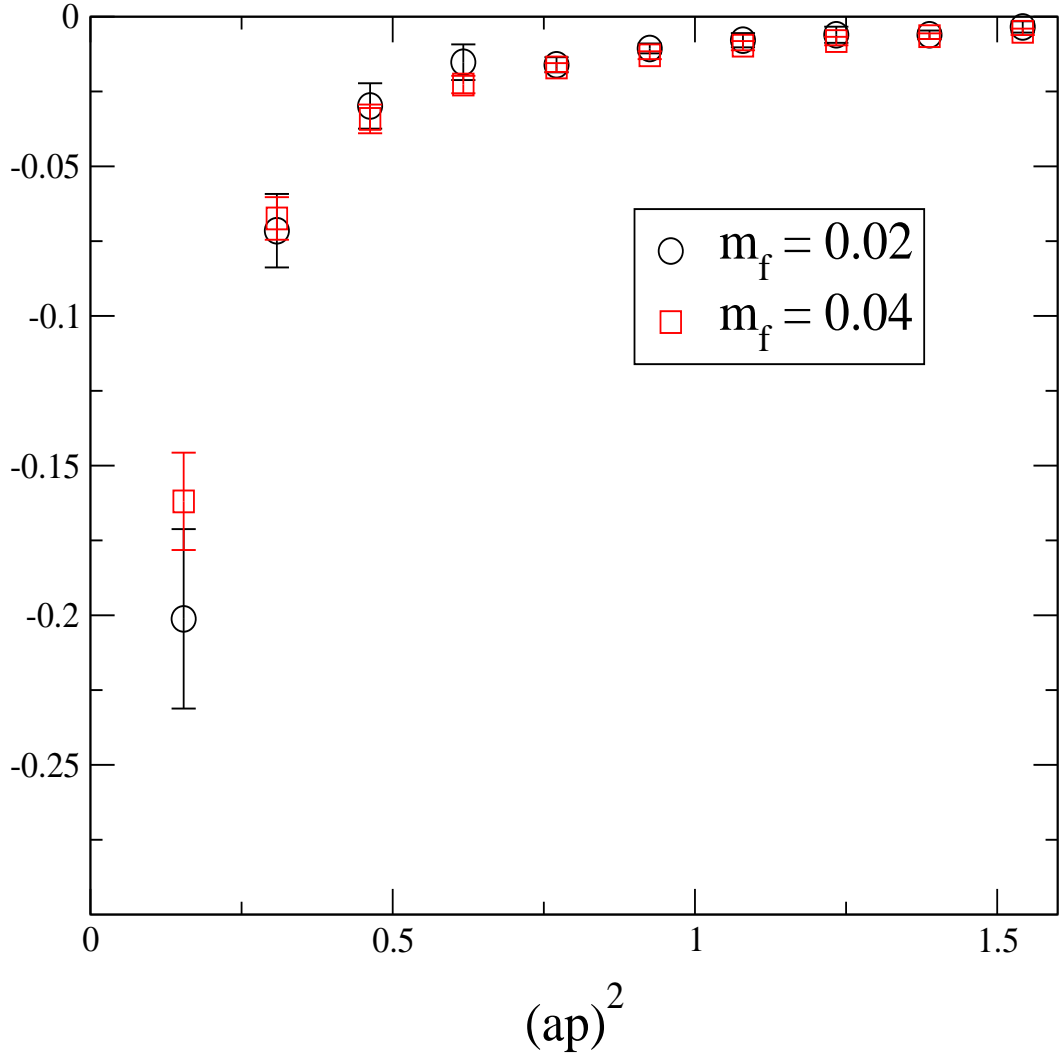}}
\caption{The difference between $Z_q$ as extracted from the conserved axial and vector currents.} 
\label{fig:vadiff}
\end{center}
\end{figure}

\begin{figure}[!htb]
\begin{center}
\epsfxsize=\hsize
\mbox{\epsfig{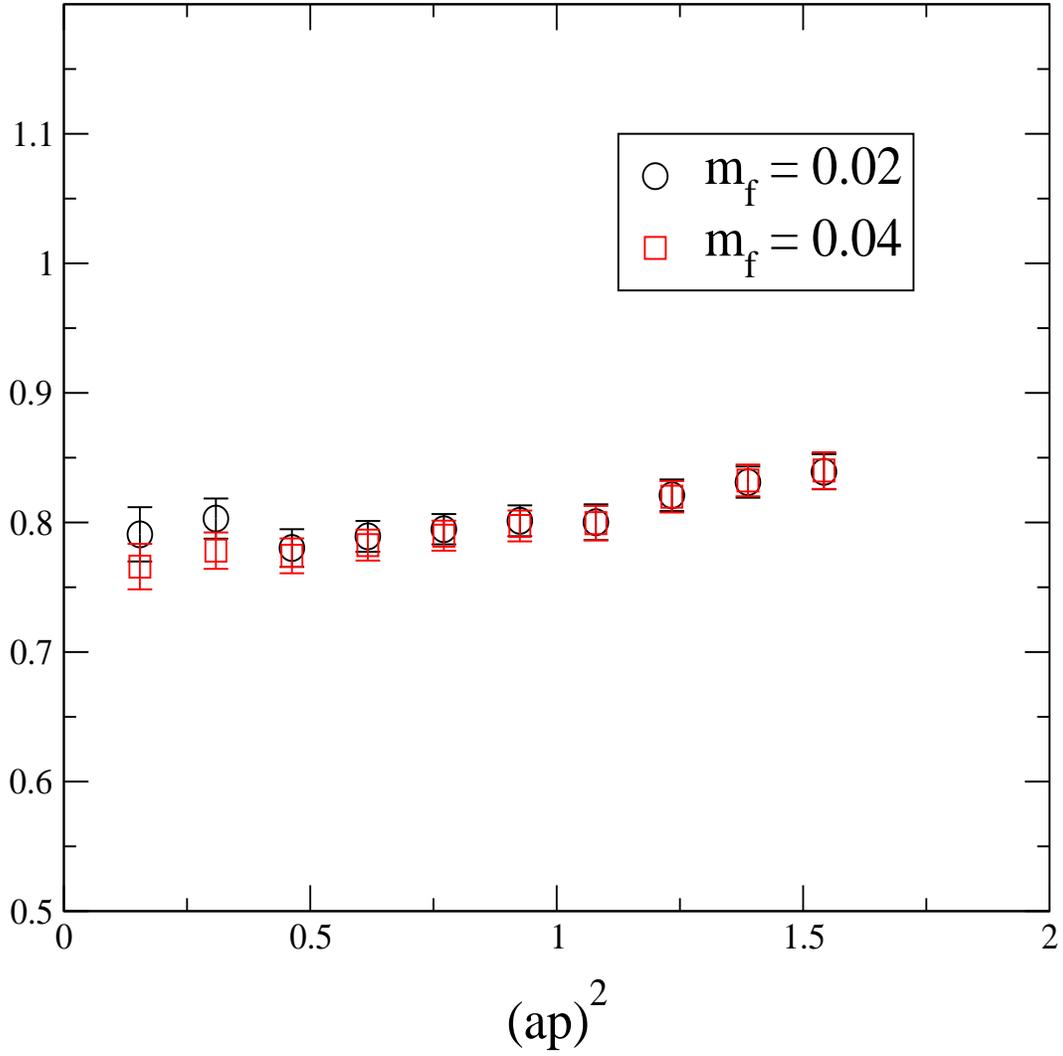}}
\caption{The average of $Z_q$  from the conserved axial and vector currents.} 
\label{fig:vaavv}
\end{center}
\end{figure}

\Fig{fig:vadiff} shows the difference between $Z_q$ calculated using the axial
and vector Ward-Takahashi identities, while \Fig{fig:vaavv} shows the average.
As can be seen from \Fig{fig:vadiff}, while for low momenta the 
two methods give different results, this difference is damped at large momenta
as would be expected if this effect stems from spontaneous (rather than
explicit) chiral symmetry breaking. Again, this provides strong
evidence that the effects of explicit chiral symmetry breaking are
negligible in these calculations.

\section{$Z_A$ from Hadronic Matrix Elements}
\label{sect:spect}

As mentioned in Section~\ref{sec:zqward}, to a good approximation the axial
current defined by \Eq{dwf:acur} is conserved, and therefore has a
renormalisation coefficient equal to the identity. This provides a simple way
to calculate the renormalisation coefficient of the local axial current
operator, $Z_A$, directly from hadronic matrix element calculations. One
method, that has been used in Ref.~\cite{Blum:2000kn}, is to note that the matrix element of
any operator with the renormalised axial current is a well defined
quantity, which will be independent of the interpolating operator for 
the axial current at distances above the scale of the lattice spacing.
Therefore, 
\be
\langle {\cal A }_{0}(t_1) \, \, \overline{q} \gamma_5 q (t_2) \rangle
 = Z_A \langle A_{0}(t_1) \, \, \overline{q} \gamma_5 q (t_2) \rangle
\label{spect:rel} \, .
\ee
for $|t_1-t_2|a \gg 1$.
A full discussion of this method and the results are given in
Ref.~\cite{Blum:2000kn}, but it is useful to summarise the results
here. Table~\ref{tab:specta} collects together values for
$Z_A$ for several $L_S$ values on a $\beta=6.0$, $16^3 \times 32$
lattice with $M_5=1.8$.
The quoted errors are statistical only.

\section{Renormalisation Group Behaviour} 
\label{sec:rg}
The previous sections have provided an extraction of the renormalisation
coefficients of interest taking into account the possible effects stemming
from chiral symmetry breaking (either explicit or spontaneous). In general,
perturbation theory predicts that these coefficients may be logarithmically
dependent on the momentum scale. Lattice artifacts may also cause the
result to depend on the definition of momentum on the lattice.  When 
small these lattice artifacts will be manifest in the data as added terms 
proportional to $\left( ap \right)^2$. 

The simplest approach to using the renormalisation coefficients calculated
in this paper is to take the (mass-pole subtracted) value
for $\Lambda_i\left(ap\right)$ as $\left(Z_q/Z_i\right)\left(ap\right)$.
However, we need to address two significant uncertainties:

First, this choice assumes, without verification, that the \order{$(ap)^2$} 
lattice artifacts are small.  We should attempt to understand the
momentum dependence of the amplitude $\Lambda_i\left(ap\right)$ to 
determine the size of the \order{$(ap)^2$} contamination and, if possible, 
remove it.

Second, most operators of interest in lattice calculations are ultimately 
defined using continuum perturbation theory.  Making this connection requires
the use of perturbation theory to connect the RI scheme and momentum scale
used here with the renormalisation procedure and momentum scale used in the 
original continuum definition, typically the $\overline{MS}$ scheme.  This 
final matching step between the normalised lattice and continuum operators 
is done at a specific momentum scale for the renormalised lattice operator.  
In general, both the normalisation of lattice operators and the matching 
coefficients will depend on this momentum scale.  It is not known 
{\it a priori} how many loops in perturbation theory must be calculated 
to correctly describe the momentum range probed in current lattice 
calculations, or even if perturbation theory can describe the 
region we are studying.  Comparing the momentum behaviour predicted from 
perturbation theory to that of the data therefore provides an important 
consistency check for the general framework of the method.

Our approach to comparing the known perturbative running of the quantities 
of interest to our numerical data will be to divide the data by the 
predicted renormalisation group running, with the overall normalisation 
set by requiring that at the point $(ap)^2 = 1$ this divisor is one.  If 
the perturbative result correctly describes the data, and the effect of 
$(ap)^2$ terms may be neglected, the result will be completely scale 
independent.

There are three components that are needed to calculate these
quantities:
\begin{enumerate}
\item The anomalous dimensions for the operators from perturbation theory.
\item The ratio of $Z^{{\rm RI}}/Z^{\overline{{\rm MS}}}$ must be known to
perform the matching.  Since the renormalisation condition 
that determines $Z^{\rm RI}$ is well defined both on the lattice and
when in continuum dimensional regularisation this ratio may be calculated
perturbatively using the latter regularisation. In general it can be
expanded as
\be
\frac{Z^{{\rm RI}}}{Z^{{\rm \overline{MS}}}}
=
1
+
\frac{\alpha_s}{4 \pi}
Z^{(1){\rm RI}}_0
+
\frac{\alpha_s^2}{\left(4 \pi\right)^2}
Z^{(2)({\rm RI})}_0 +
\ldots \, .
\label{rg:match}
\ee
For a consistent treatment this ratio need only be known to one less
power of $\alpha_s$ than the running is known.
\item A lattice value for $\alpha_s$. The value of $\alpha_s$ affects the scale
dependence of both the matching and running for this calculation. For this
work the value of $\alpha_s$ was calculated at three loops using 
a lattice value of $\Lambda_{QCD}$
taken from Ref.~\cite{Capitani:1998mq}
as
\be \Lambda_{QCD} = 238 \pm 19 \mathrm{MeV} \, .  \ee 
To do this consistently with the way the lattice treatment
in Ref.~\cite{Capitani:1998mq} was performed, their value of $r_0 = 0.5 \,
\mathrm{fm}$ was taken and converted into a lattice spacing using the results
of Ref.~\cite{Guagnelli:1998ud}.  For the dimensionful scales that we will quote,
we set the physical scale through the rho mass computed with domain wall
fermions \cite{Blum:2000kn}, which for $\beta=6.0$ gives
\be
a = 0.520(11) \mathrm{GeV}^{-1}\, .
\ee
\end{enumerate}

\begin{figure}[!htb]
\begin{center}
\epsfxsize=\hsize
\mbox{\epsfig{file=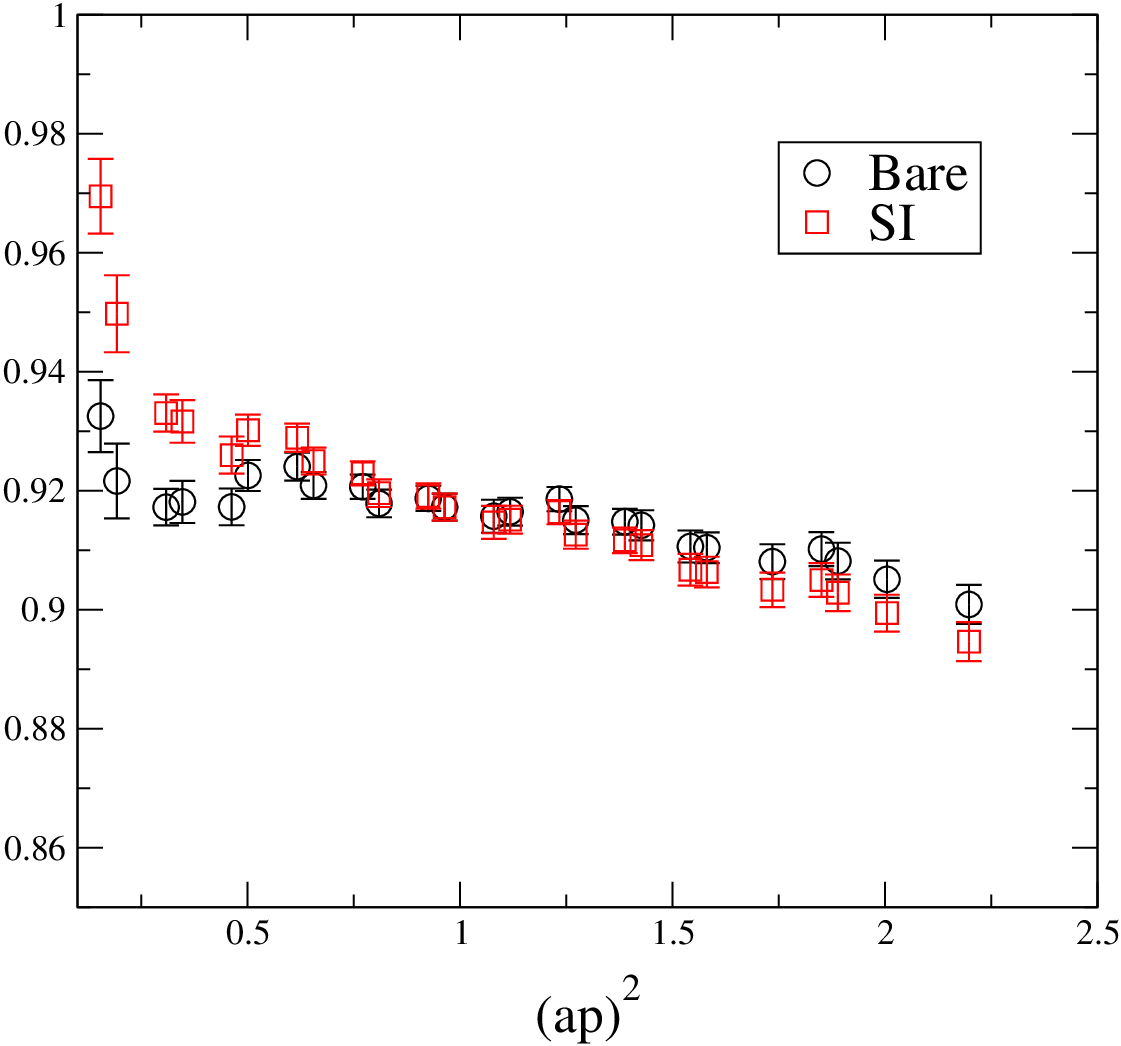,width=14truecm, height=14truecm}}
\caption{A plot showing the raw data for
 $1/\Lambda_A$ (labeled as ``Bare'') and the 
value of  $1/\Lambda_A$ divided by its predicted three loop
perturbative running (labeled as ``SI''), such that they coincide at
$(ap)^2=1$, versus momentum. 
The slope of the latter versus $(ap)^2$ may be interpreted as 
an \order{$a^2$} effect and is $\approx -0.02$.} 
\label{fig:zarg}
\end{center}
\end{figure}

\begin{figure}[!htb]
\begin{center}
\epsfxsize=\hsize
\mbox{\epsfig{file=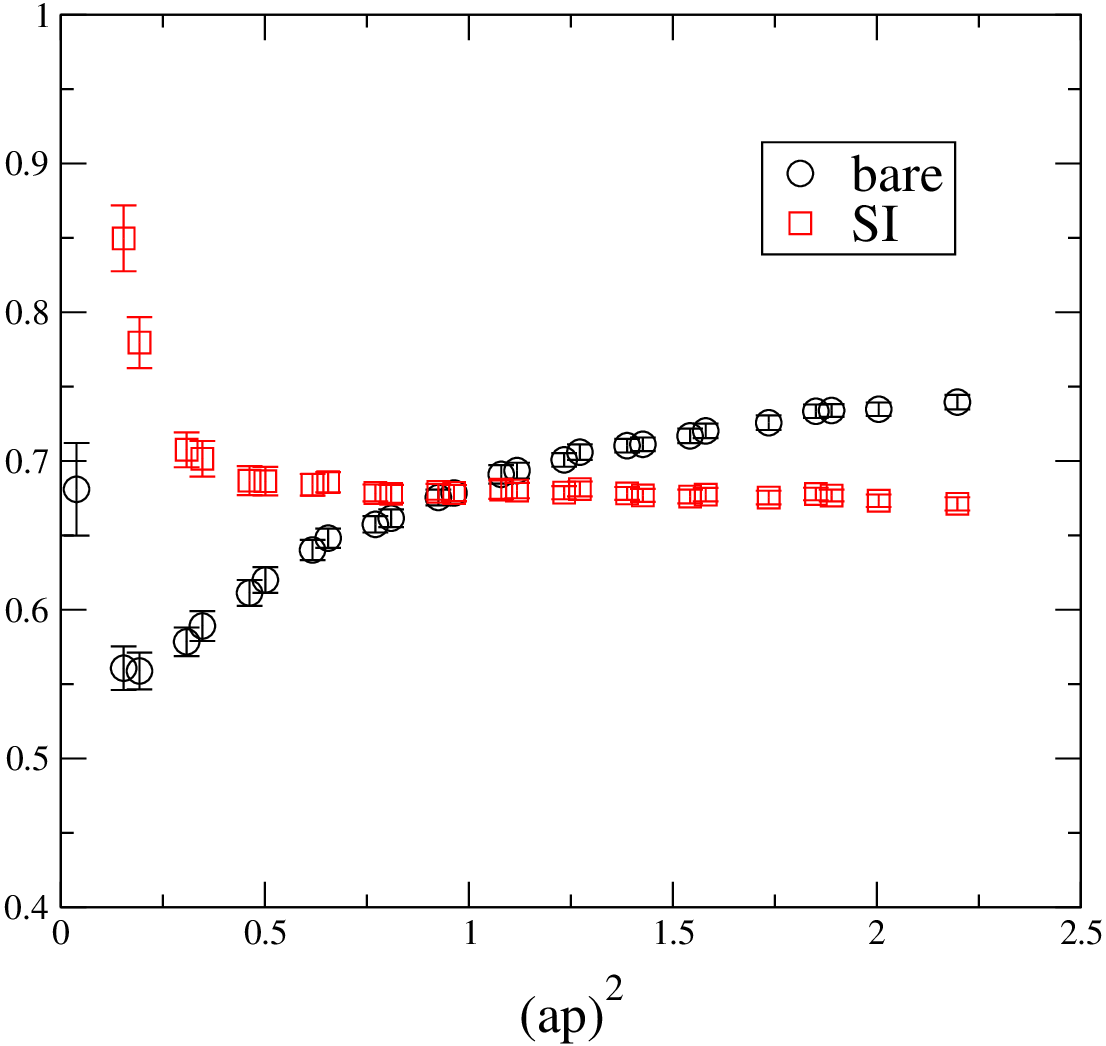,width=14truecm, height=14truecm}}
\caption{A plot showing the raw data for
 $1/\Lambda_S$ (labeled as ``Bare'') and the 
value of  $1/\Lambda_S$ divided by its predicted three loop
perturbative running (labeled as ``SI''), such that they coincide at
$(ap)^2=1$, versus momentum. 
The slope of the latter versus $(ap)^2$ may be interpreted as 
an \order{$a^2$} effect and is $\approx -0.003$.} 
\label{fig:zsrg}
\end{center}
\end{figure}

\begin{figure}[!htb]
\begin{center}
\epsfxsize=\hsize
\mbox{\epsfig{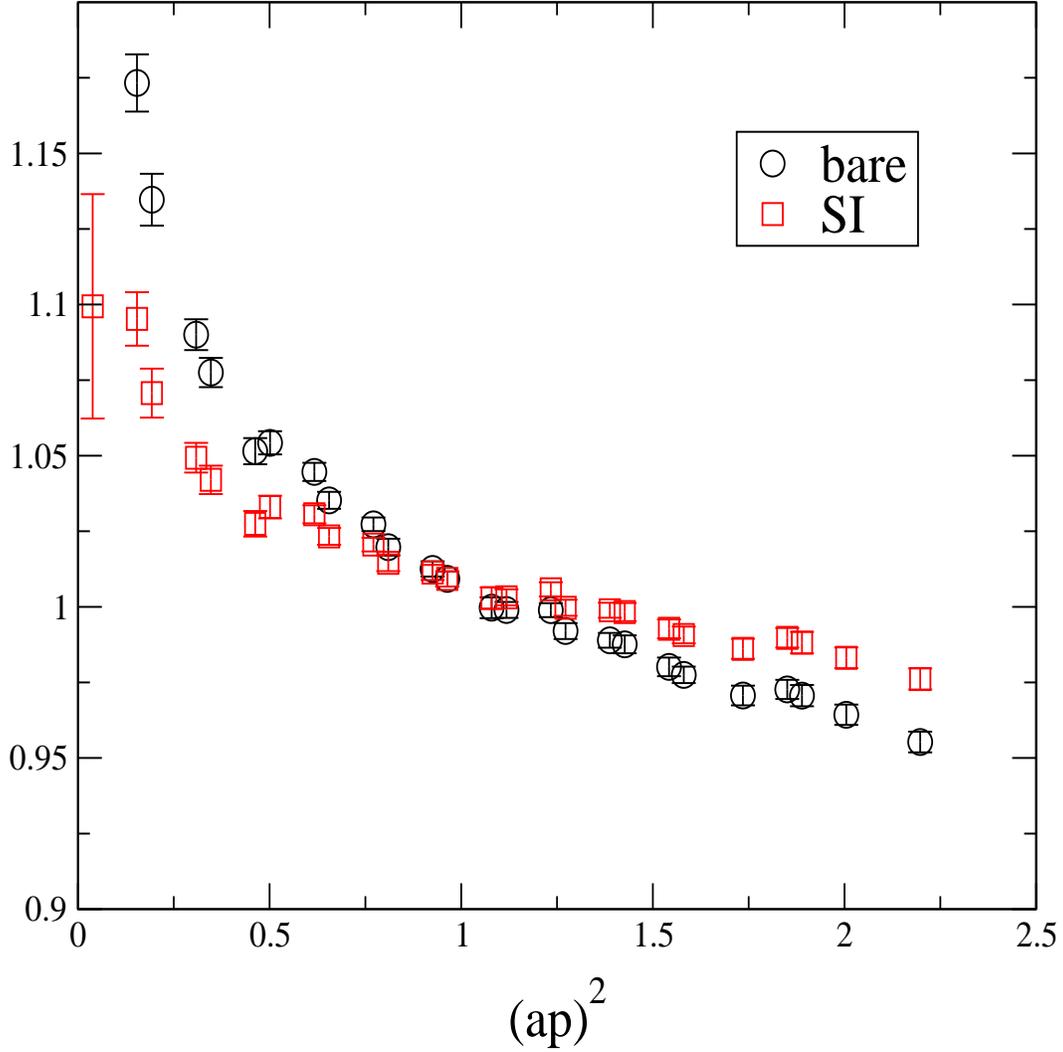}}
\caption{A plot showing the raw data for
 $1/\Lambda_T$ (labeled as ``Bare'') and the 
value of  $1/\Lambda_T$ divided by its predicted one loop
perturbative running (labeled as ``SI''), such that they coincide at
$(ap)^2=1$, versus momentum. 
The slope of the latter versus $(ap)^2$ may be interpreted as 
an \order{$a^2$} effect and is $\approx -0.02$.} 
\label{fig:zrrg}
\end{center}
\end{figure}

Both $Z_A$ and $Z_V$ should be scale independent, but this is not the case for
$Z_q$. \Fig{fig:zarg} shows both $1/\Lambda_A$ and the scale invariant (SI)
quantity calculated as described above:
\be
\Lambda_A^{SI}\left((ap)^2\right) = \Lambda_A\left((ap)^2\right)/C_A\left((ap)^2\right).
\ee
The quantity $C_A$ is determined through three loops using the anomalous
dimension coefficients calculated in 
Refs.~\cite{Chetyrkin:1999pq,Gimenez:1998ue,Franco:1998bm} as described in
Appendix~\ref{sec:Z_factors}.  It is normalised so that $C_A(1)=1$.
As can be seen, in
this case the renormalisation group running actually goes in the opposite
direction from the data. The scale dependence of this data, either predicted
or actual, is, however, very small and a plausible explanation for this is an
$(ap)^2$ error. Indeed, when a linear fit of the SI data versus $(ap)^2$ is
performed, for $0.8 < (ap)^2 < 2.0$, the gradient is $\approx -0.02$. 
A more compelling test of the renormalisation group behaviour is provided by
studying the data for $Z_S/Z_q$. In this case the predicted scaling behaviour
over the range of momenta studied is much larger and, as
\Fig{fig:zsrg} shows, the agreement between the predicted behaviour and the
data is impressive (with a gradient, in this case, of $\approx -0.003$).
The values for $Z_S/Z_q$ versus momentum used here are taken
after the mass-pole has been subtracted and, again, the three loop
results for the running are taken from Refs.~\cite{Chetyrkin:1999pq,Gimenez:1998ue,Franco:1998bm}.
Unfortunately, a matching calculation for $Z_T$ could not be found in the
literature, so  the data could only be compared to 
the  one loop running (which is taken from Ref.~\cite{He:1995gz}). 
The SI quantity so calculated is shown in \Fig{fig:zrrg} and has
a gradient of $\approx -0.02$.

\begin{figure}[!htb]
\begin{center}
\epsfxsize=\hsize
\mbox{\epsfig{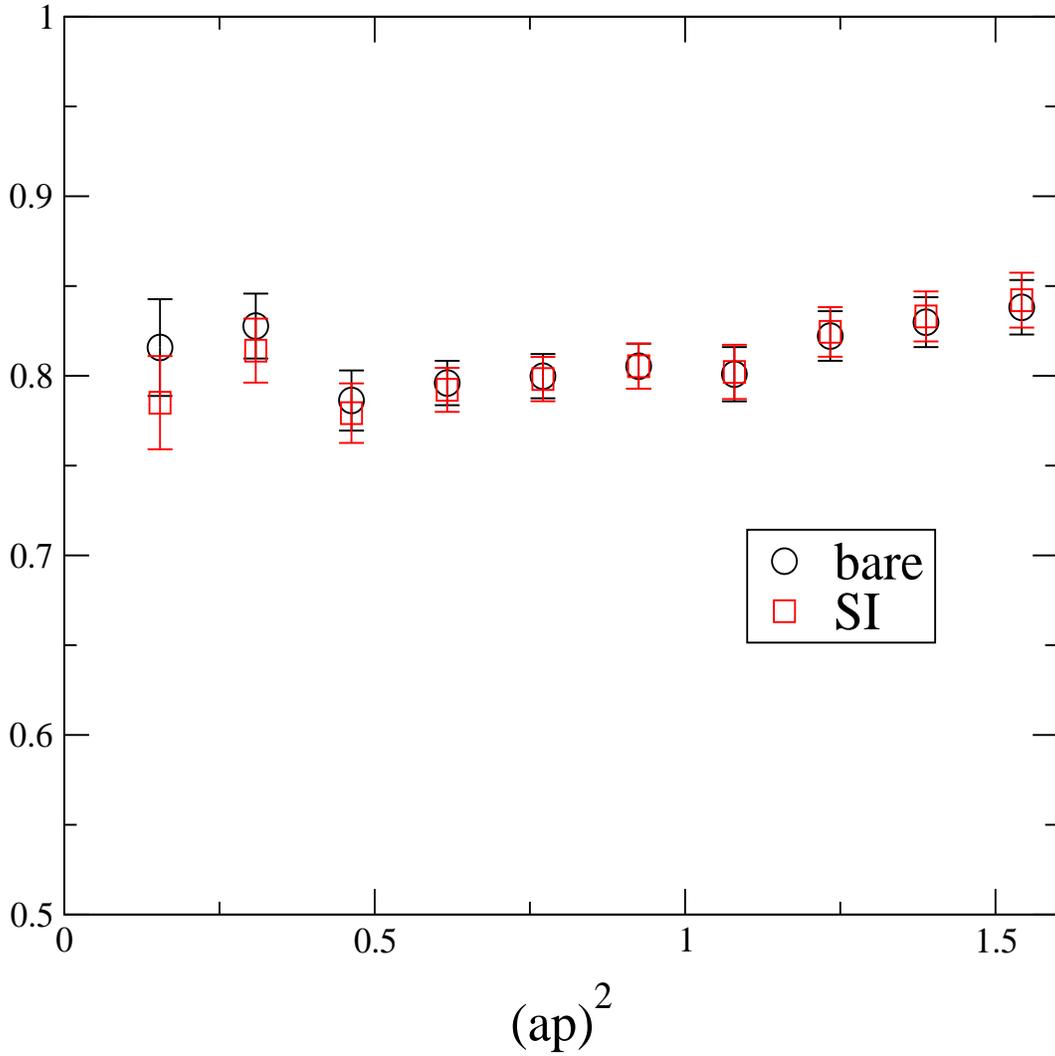}}
\caption{Bare and scale invariant (SI) versions of $Z_q$ determined from 
the conserved axial and vector currents.}
\label{fig:zqrgi}
\end{center}
\end{figure}

Taking the interpretation that the remaining scale dependence is due
to \order{$(ap)^2$} effects, the correct way to extract the renormalisation
coefficients is to first construct the SI quantity as described above, and
then fit any remaining scale dependence~\cite{Gimenez:1998ue} to  
the form
\be
y = c_1 + c_2  (ap)^2 \, ,
\ee
for a range of momenta that is chosen to be ``above'' the region for
which condensate effects are deemed to be important. Table~\ref{tab} shows
the fitted values for the RI and $\overline{MS}$ scheme renormalisation
coefficients using a fitting rage of $ 0.8 < (ap)^2 < 2.0$.
Now that the renormalisation group running has been taken into account, it is
possible to make a comparison of the various methods of calculating $Z_q$ and
thus give final results for the renormalisation factors. Table~\ref{tab}
already gives $Z_q$ as calculated from the conserved currents (\Fig{fig:zqrgi}
shows the momentum dependence of both the SI and bare form).
Another simple way to derive this quantity is by taking $Z_A/Z_q$ from Table~\ref{tab}
and combining it with the value of $Z_A$ obtained from hadronic matrix elements.
This gives $Z_q = 0.805(3)(15)$. This, approximately $5\%$, difference may
be taken as an indication of the size of the systematic errors.
%

\section{$Z_q$ from the Propagator -- Results}
\label{sect:zq_prop}

Here we show results for the wave function
renormalisation computed through \Eq{num:zqprime}
and demonstrate that this methods contains a comparatively
large systematic uncertainty due to the ambiguity
in defining discrete momentum.
We use the perturbative matching between ${Z'}_q$ and $Z_q$, as
given in Ref.~\cite{Franco:1998bm}.\footnote{Their convention
is that a given $Z$-factor in Ref.~\cite{Franco:1998bm} 
is the reciprocal of ours.}  Then the SI $Z_q$ can be 
constructed as described above.  

\begin{figure}[!htb]
\begin{center}
\epsfxsize=\hsize
\mbox{\epsfig{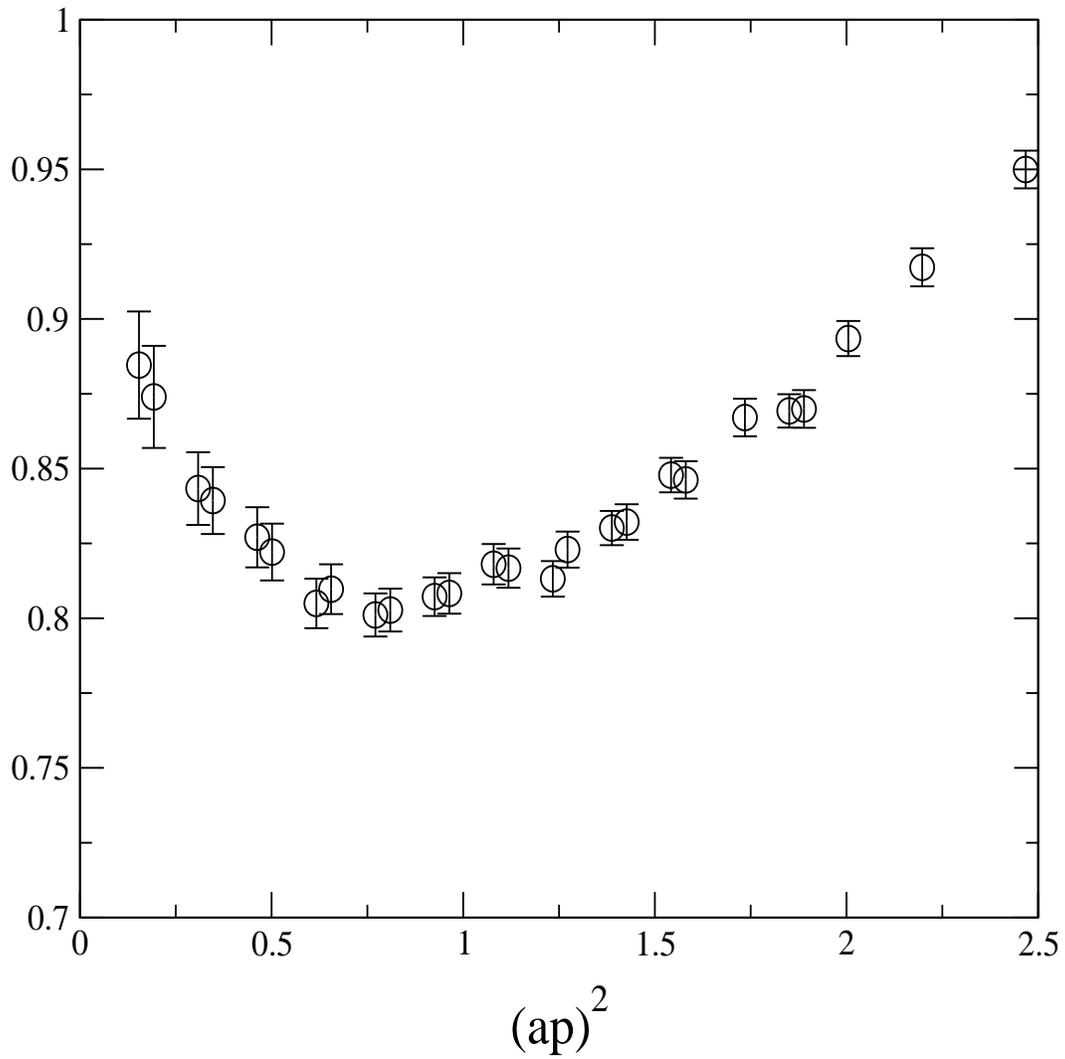}}
\caption{A scale invariant (SI) version of $Z_q$ determined from ${Z'}_q$.}
\label{fig:zqp}
\end{center}
\end{figure}

\begin{figure}[!htb]
\begin{center}
\epsfxsize=\hsize
\mbox{\epsfig{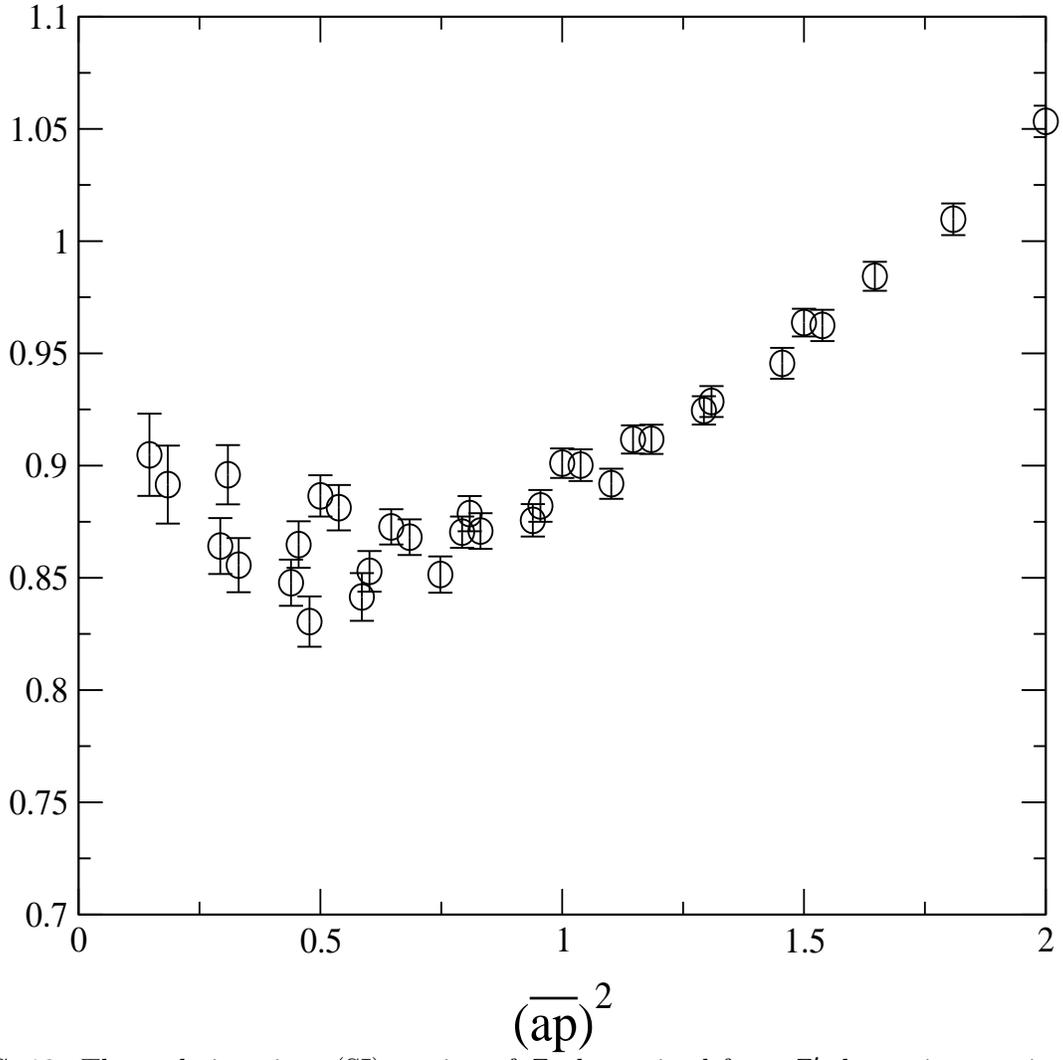}}
\caption{The scale invariant (SI) version of $Z_q$ determined from 
${Z'}_q$ but using $a\bar{p}_\mu$ instead of $ap_\mu$.}
\label{fig:zqpbar}
\end{center}
\end{figure}

\Fig{fig:zqp} shows the SI $Z_q$
using ${Z'}_q$ defined in \Eq{num:zqprime}, and
\Fig{fig:zqpbar} shows the SI $Z_q$ where
the replacement 
\be
ap_\mu \to \sin(ap_\mu) \equiv a\bar{p}_\mu
\ee
is made in \Eq{num:zqprime}.  Note that the former
is plotted vs.\ $(ap)^2$ and the latter vs.\ $(a\bar{p})^2$.
We use the data at $m_f = 0.02$ since no mass dependence can
be observed.

As in the previous section, we extrapolate to $(ap)^2 = 0$.
We find for the data in \Fig{fig:zqp}, 
$Z_q = 0.715 \pm 0.007 \pm 0.040$, where the first error
is statistical and the second comes from different choices
for the range of momenta over which to fit.  The data in
\Fig{fig:zqpbar} give $Z_q = 0.733 \pm 0.007 \pm 0.050$.
We can further probe these discretisation uncertainties
by extrapolating $Z_q^{-1}$ to zero $(ap)^2$ or $(a\bar{p})^2$.
This results in $Z_q = 0.732 \pm 0.006 \pm 0.020$
and $Z_q = 0.772 \pm 0.006 \pm 0.020$, respectively.

The spread in values of $Z_q$ obtained depending on momentum
ranges and on the definition of discrete momenta mean that 
extracting a SI $Z_q$ in the same manner as the $Z$'s for
bilinear operators is less precise.  However, the results are in rough
agreement with the more precise methods described above.

\section{Comparison with Perturbation Theory}
\label{sect:pert}

All the renormalisation factors considered above have also been calculated in
lattice perturbation theory, at the one loop level, for the domain wall
fermion action in the $L_s \rightarrow \infty$ limit
\cite{Aoki:1997xg,Aoki:1998vv,Blum:1999xi}.  As we see little evidence of
explicit chiral symmetry breaking effects in our study, the fact that the
perturbative calculations have been 
performed in the $L_s \rightarrow \infty$ limit
will probably not affect this comparison. However, a more serious issue is
which $M_5$ value to use in the perturbative formulae. 

The reason for this is easy to understand. Away from the walls, the massless
domain wall fermion Lagrangian, \Eq{dwf:D_DWF}, may be viewed as a simple
extension of the standard Wilson fermion action to five dimensions with a
negative mass term, $M_5$. Like the four dimensional Wilson mass term, $M_5$
undergoes a strong additive renormalisation, the size of which perturbation
theory is not good at predicting. While a more accurate prediction may be made
using tadpole improved perturbation theory \cite{Lepage:1993xa}, a good
deal of ambiguity remains in the perturbative prediction of any quantity that
is strongly dependent on $M_5$.

Further consideration of the similarity of the domain wall and standard Wilson
actions, leads to a non-perturbative estimate of the magnitude of this
additive renormalisation.  As argued in Ref.~\cite{Blum:1999xi}, in the $L_s
\rightarrow \infty$ limit, the effect of this additive renormalisation
may be taken into account by using 
\be
M_5^{\mathrm{pert}} 
=  
M_5^{\mathrm{sim}} - \left(4 - \frac{1}{2\kappa_c}\right) \, ,
\label{eq:m5}
\ee
in the perturbative equations,
where $\kappa_c$ is the four-dimensional critical Wilson hopping parameter
and $M_5^{\mathrm{sim}}$ refers to the value used in the non-perturbative
simulation (in our case $M_5^{\mathrm{sim}}$=1.8).
For $\beta =6.0$, this ansatz leads to
shift of magnitude $\approx0.8$, which has been
found to describe the dependence of the 
pion mass squared as a function of $M_5$ to a good
degree of accuracy in a numerical simulation with
$L_s=14$\cite{Blum:1999xi}.  

Eq. 4.10 in Ref.~\cite{Aoki:1998vv} gives
the complete one loop bilinear renormalisation constants in the
$\overline{MS}$ scheme:
\bea
Z_{\Gamma}^{\rm total} &=& ( (1-w_0^2) Z_w )^{-1} Z_{\Gamma}.
\label{eq:Z_PT}
\eea Here $Z_w$, $Z_2$, and $Z_{\Gamma}$ are to be computed from Eqs. 3.30,
3.42, and 4.11 and Tables 2 and 3 in Ref.~\cite{Aoki:1998vv}, while $w_0 =
1-M_5$.  In the mean-field improved case, the above relations hold with the
replacements $w_0\to w_0^{MF}=w_0+4(1-u)$, $Z_w\to Z_w^{MF}$, $Z_{2}\to u
Z_{2}^{MF}$, and $Z_{\Gamma}\to u Z_{\Gamma}^{MF}$ \cite{aoki:priv}, whose
values can also be computed from Tables 2 and 3 in
Ref.~\cite{Aoki:1998vv}. The factor $u$ in these formulae is the mean link
variable in Feynman gauge. As it is not possible to use the value of the mean
link in Feyman gauge we have instead used the fourth root of the plaquette and
the perturbative results of Ref.~\cite{Aoki:1999ky} to convert the results of
Ref.~\cite{Aoki:1998vv}.

\begin{figure}[!htb]
\begin{center}
\epsfxsize=\hsize
\mbox{\epsfig{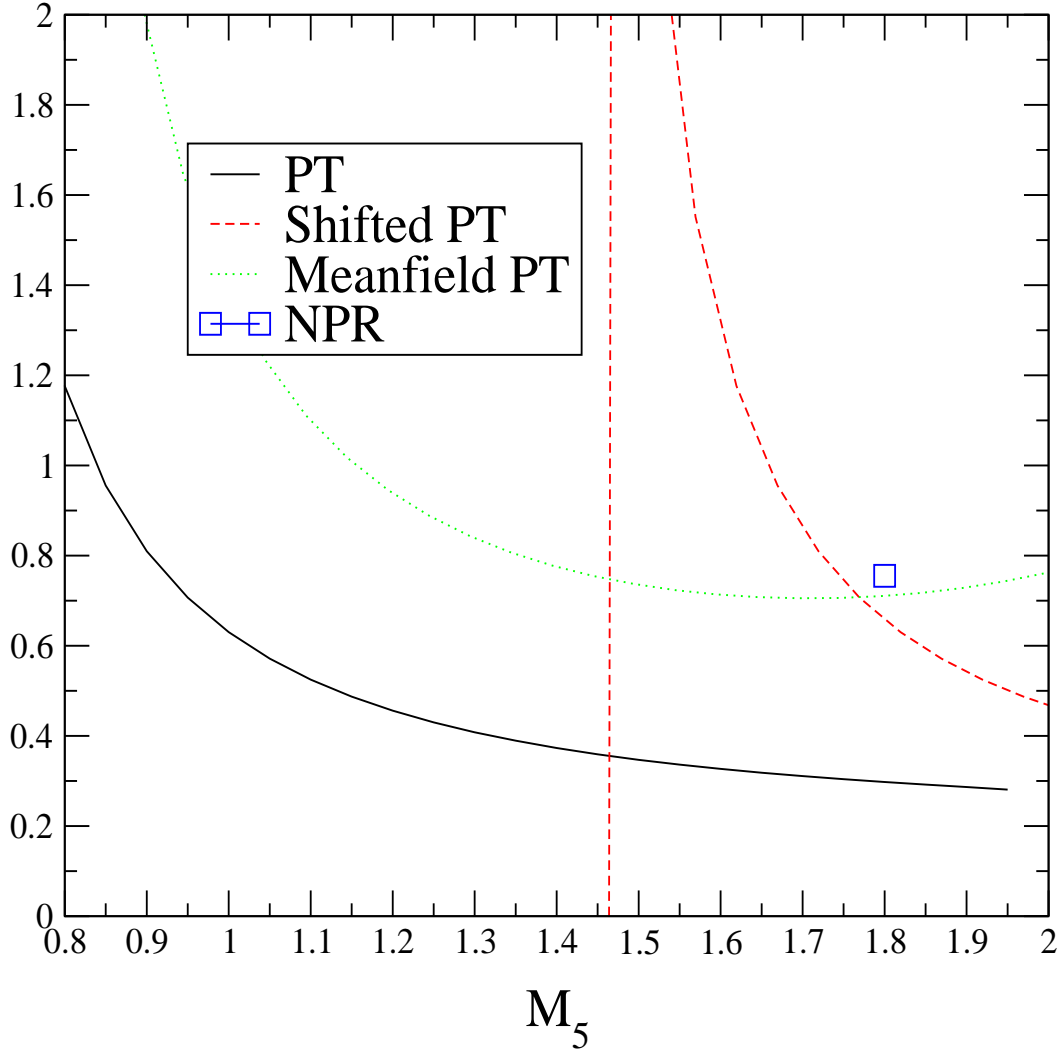}}
\caption{The renormalisation factor $Z_A^{\rm total}$ of 
Eq.~\protect{\ref{eq:Z_PT}} in the $\overline{\rm MS}$ scheme at 
$2~\mathrm{GeV}$ computed in naive perturbation theory, naive 
perturbation theory shifted by $(4-1/2\kappa_c)$ as in 
Eq.~\protect{\ref{eq:m5}} and mean field improved perturbation theory.} 
\label{za}
\end{center}
\end{figure}

\begin{figure}[!htb]
\begin{center}
\epsfxsize=\hsize
\mbox{\epsfig{file=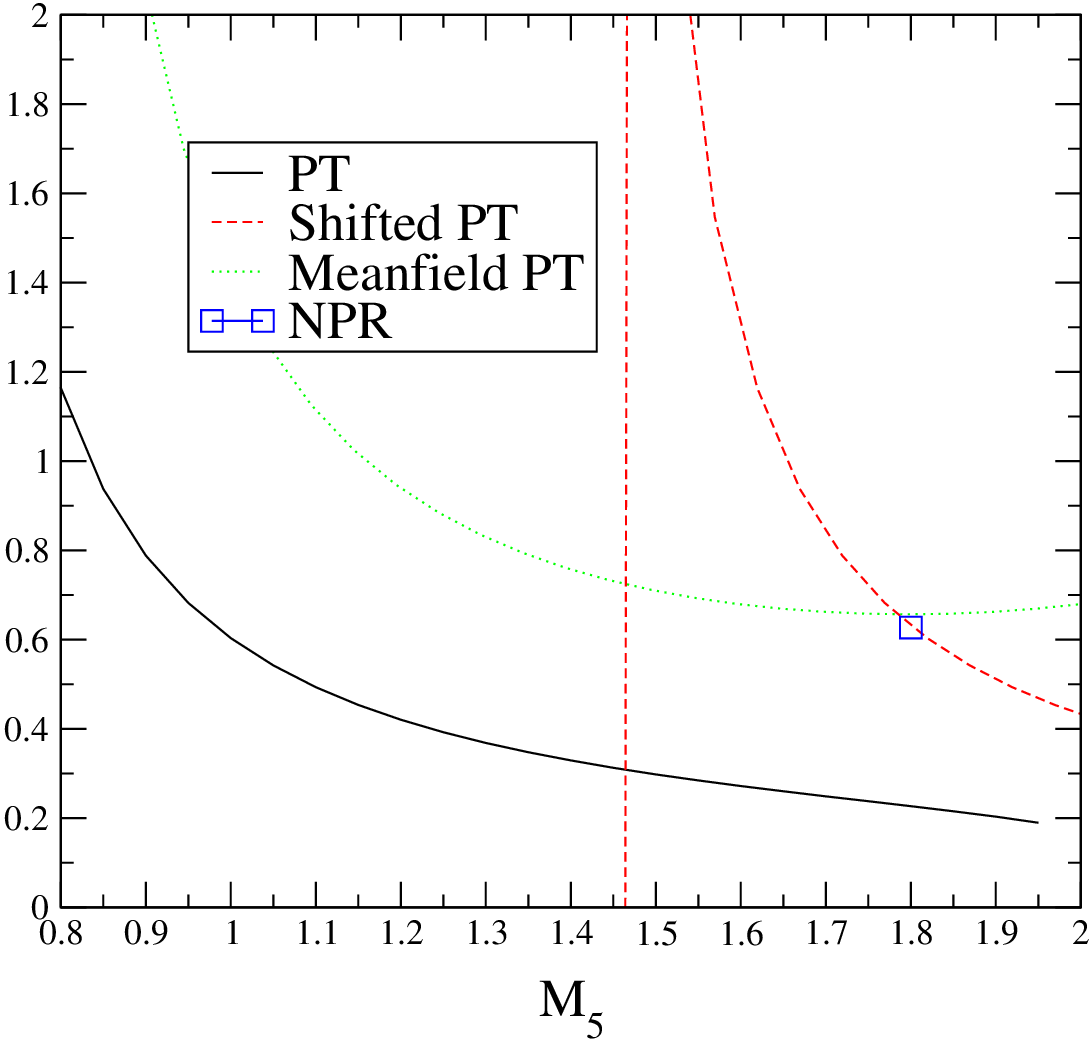,width=14truecm, height=14truecm}}
\caption{Same as \Fig{za} but for $Z_S^{\rm total}$.}
\label{zs}
\end{center}
\end{figure}

In \Fig{za} and \Fig{zs} we plot $Z^{\rm total}_A$ and 
$Z^{\rm total}_S$, respectively,
as functions of the variable $M_5$ in naive perturbation theory, in
naive perturbation theory with the variable $M_5$ shifted 
according to Eq.~\ref{eq:m5} and in the mean-field improved case.
To compute $\alpha_s$, 
we used the same input values for $\Lambda_{QCD}$ and $a$ as
in the perturbative running calculations in \Sect{sec:rg}.
We obtain $\alpha_s\left( (ap)^2 = 1 \right)= 0.20$.
These figures show appreciable $M_5$ dependence.
Our non-perturbative result is shown as a point corresponding to the
single value of $M_5=1.8$ that we have studied.

The naive perturbation theory curve has a significant dependence
on the precise value of $\alpha_s$. In the mean-field improved case this
problem is not as serious as the coefficient of $\alpha_s$ is a factor of
2-3 times smaller.   Examining Figures \ref{za} and \ref{zs}, one recognises 
that naive perturbation theory does a poor job of determining $Z_A$ or $Z_S$
giving values nearly 2 times too small for $M_5=1.8$.  Introducing the
shift of Eq.~\ref{eq:m5} improves the situation noticeably giving 
values 15\% too small and to within a few percent, although the
perturbative result is rapidly varying with $M_5$ in this case.  
The mean-field results differ from the non-perturbative
result by around 5\% in both cases.

\section{Conclusions}

In this paper we have described a first study of
non-perturbative renormalisation
of the quark field and flavour non-singlet
fermion bilinear operators in the context of domain wall fermions.
We presented a theoretical argument constraining the form that
explicit chiral symmetry breaking effects may take, and 
found that numerically these are insignificant, as might be expected
from the measured size of the additive mass renormalisation,
$m_{\mathrm{res}}$, \cite{Blum:2000kn,AliKhan:2000iv}.
However, systematic effects due to spontaneous chiral symmetry breaking 
and zero-modes are significant, but accurately follow the expected 
form and can be effectively subtracted away.

Renormalisation group invariant quantities were obtained in
\Sect{sec:rg} by dividing the regularisation independent
scheme coefficients by the three loop renormalisation group 
running (where available). The residual scale dependence of
these quantities is small and was treated as an \order{$a^2$}
error. Three different quantities were used to determine the quark 
renormalisation factor: the off-shell vertex functions of the conserved 
vector and axial currents; the trace of the product of $p^\mu\gamma^\mu$ 
and the off-shell quark propagator; and the combination of $Z_A$ as 
determined from hadronic matrix elements with the value of $Z_A/Z_q$ 
obtained in this study from the off-shell, axial vector vertex function. 
The technique of obtaining this $Z_q$ directly from the propagator suffers from
large discretisation errors, but is roughly consistent with the 
other two methods which gave results differing by $\approx 5\%$.

In the final section we compared our results against the predictions
of both standard and mean-field improved one loop perturbation theory. 

\section*{Acknowledgments}

The authors would like to acknowledge useful discussions
with Sinya Aoki.
We thank RIKEN, 
Brookhaven National Laboratory and the U.S. Department of Energy for 
providing the facilities essential for the completion of this work.

This research was supported in part by the DOE under grant \#
DE-FG02-92ER40699 (Columbia), by the NSF under grant \#
NSF-PHY96-05199 (Vranas), by the DOE under grant \#
DE-AC02-98CH10886 (Dawson-Soni) and  by the RIKEN-BNL Research
Center (Blum-Wingate).

\appendix
\section{The Running of $\alpha_s$}
In the following the definitions
\be
\alpha = \frac{g^2}{4 \pi} \, ,
\ee
and
\be
C_F = \frac{N_C^2 -1}{2 N_C} \, ,
\ee
will be used.
The renormalised coupling may be defined in terms of the bare coupling by,
\be
\alpha_b = Z_g^2 \alpha_s \mu^{2 \epsilon} \, .
\ee
As $\alpha_b$ is completely independent of $\mu$,
\be
\mu^2 \frac{d \alpha_s }{d \mu^2}
=
- \epsilon \alpha_s
+ \beta  
\ee
with,
\be
\beta = - \alpha_s \frac{2}{Z_g}\mu^2 \frac{ d Z_g}{d \mu^2}
\ee 
The results for the beta function are most easily given in
terms of the $\beta_i$ variable:
\be
\frac{\beta(\alpha_s)}{4 \pi} = 
- \beta_0 \left[ \frac{ \alpha_s}{ 4\pi } \right]^2
- \beta_1 \left[ \frac{ \alpha_s}{ 4\pi } \right]^3
- \ldots \, .
\ee
The values used in the current work are summarised in
Table~\protect{\ref{tab:bet}}. They are taken from Ref.~\cite{Franco:1998bm} with
the number of flavours set to zero (as we are working in the quenched case) 
and the number of colours set to three.

Once the $\beta_i$'s are known the running equation,
\be
\mu \frac{d}{d \mu} \alpha_s
=
-2 \beta_0 \frac{\alpha_s^2}{ 4 \pi }
-2 \beta_1 \frac{\alpha_s^3}{ (4 \pi)^2 }
- \ldots \, ,
\ee 
may be solved:
\begin{itemize}
\item 
One loop solution:
\be
\frac{\alpha_s}{4 \pi} = \frac{1}{\beta_0 \ln \left( \mu^2 / \Lambda^2_{QCD} \right) }
\ee
\item
Two loop solution \cite{Buras:1998ra}:
\be
\frac{\alpha_s}{4 \pi}
=
\frac{1}{\beta_0 \ln \left( \mu^2 / \Lambda^2_{QCD} \right) }
-
\frac{\beta_1 \ln \ln \left( \mu^2 / \Lambda^2_{QCD} \right)
}{
\beta_o^3 \ln^2 \left( \mu^2 / \Lambda^2_{QCD} \right)}
\ee
\item
Three loop solution \cite{Gimenez:1998ue}:
\bea
\frac{\alpha_s}{4 \pi}
&=&
\frac{1}{\beta_0 \ln \left( \mu^2 / \Lambda^2_{QCD} \right) }
-
\frac{\beta_1 \ln \ln \left( \mu^2 / \Lambda^2_{QCD} \right)
}{
\beta_o^3 \ln^2 \left( \mu^2 / \Lambda^2_{QCD} \right)}
\\ \nonumber
&&
+
\frac{1}{\beta_0^5 \ln ^3\left( \mu^2 / \Lambda^2_{QCD} \right)
}\left\{
\beta_1^2 \ln^2 \ln \left( \mu^2 / \Lambda^2_{QCD} \right)
- \beta_1^2 \ln \ln \left( \mu^2 / \Lambda^2_{QCD} \right)
\right.
\\ \nonumber
&&
\left.
+ \beta_2 \beta_0 - \beta_1^2
\right\}
\eea
\end{itemize}
\section{The Running of the $Z$-factors}
\label{sec:Z_factors}
As mention previously, the renormalised operators we are working with are
 defined as
\be
Z_{O} O_{bare} = O_{ren} \, .
\ee
Requiring that the bare operator is independent of the renormalisation
scale gives the RG equation,
\bea
\mu \frac{d}{d \mu} O_{ren} 
&=&  \frac{1}{Z_O} \mu \frac{d Z_{O}}{d \mu}  O_{ren} \, 
\\
&=&
-\frac{\gamma_O}{2} O_{ren} \, .
\eea
Writing the solution to this equation as
\be
Z_O\left(\mu^2\right) = 
\frac{
C_O\left({\mu'}^2\right)
}{
C_O\left({\mu}^2
\right)
} Z_O\left({\mu'}^2\right) \, .
\ee
and using the notation
\bea
\gamma_O  &=& 
\sum_i \gamma^{(i)}_O \left( \frac{\alpha_s}{4 \pi}\right)^{i+1} \\
\overline{\gamma}_{Oi} &=& \frac{\gamma^{(i)}_O}{2 \beta_0} \\
\overline{\beta}_i &=& \frac{\beta_i}{\beta_0} \, ,
\eea
gives rise to solutions to the running equation of the form (where we 
have suppressed the subscripts identifying the particular operator O):
\begin{itemize}
\item One loop solution \cite{Buras:1998ra}:
\be
C(\mu^2) = \alpha_s(\mu)^{\overline{\gamma}_0}.
\ee
\item Two loop solution:
\be
C(\mu^2) =
 \alpha_s(\mu)^{\overline{\gamma}_0}
\left\{
1
+ \frac{\alpha_s(\mu)}{4 \pi} \left( \overline{\gamma}_1
-\overline{\beta}_1 \overline{\gamma}_0 \right)
\right\}.
\ee
\item Three loop solution:
\bea
C(\mu^2) &=&
 \alpha_s(\mu)^{\overline{\gamma}_0}
\left\{
1
+ \frac{\alpha_s(\mu)}{4 \pi} \left( \overline{\gamma}_1
-\overline{\beta}_1 \overline{\gamma}_0 \right)
\right. 
\\ \nonumber
&&
\left.
+
\frac{1}{2} \left( \frac{\alpha_s(\mu)}{4\pi} \right)^2
\left[
\left(
\overline{\gamma}_1 - \overline{\beta}_1 \overline{\gamma}_0
\right)^2
+ \overline{\gamma}_2
+ \overline{\beta}_1^2 \overline{\gamma}_0
- \overline{\beta}_1 \overline{\gamma}_1
- \overline{\beta}_2 \overline{\gamma}_0
\right]
\right\}.
\eea
\end{itemize}
Tables~\protect{\ref{tab:zqr}} to \protect{\ref{tab:zqt}} show the
anomalous dimensions used in this work. These values were taken from
Refs.~\cite{Chetyrkin:1999pq,Gimenez:1998ue,He:1995gz} with the number of flavours
set to zero and the number of colours to three.
\section{Matching Coefficients}
The numerical values of the 
matching coefficients, $Z^{(1)}_0$ and $Z^{(2)}_0$ in \Eq{rg:match},
used for $Z_q$ and $Z_S$ are collected together in
Table~\protect{\ref{app:matchq}} and Table~\protect{\ref{app:matchS}}.

\begin{table}[!htb]
\begin{tabular}{ccc}
        $L_S$  &    $Z_A$ &  no. configs. \\ \hline
        12  & 0.7560(3)  &     56 \\       
        16  & 0.7555(3)  &     56 \\
        24  & 0.7542(3)  &     56 \\   
        32  & 0.7535(3)  &     72 \\                
        48  & 0.7533(3)  &     64 \\ 
\end{tabular}
\caption{$Z_A$ computed from the ratios of hadronic 
matrix elements.}
\label{tab:specta}
\end{table}                         

\begin{table}[!htb]
\begin{tabular}{ccc}
Z - factor& RI/SI & $\overline{MS}$ at $2{\rm GeV}$ \\
\hline
$Z_A/Z_q$    & 0.934 (2)(10)  &  0.938 (2)(12)  \\ 
$Z_S/Z_q$    & 0.683 (7)(30)  &  0.779 (8)(35)  \\ 
$Z_T/Z_q$    & 1.034 (3)(100) &  1.035 (3)(100) \\ 
$Z_q^{\rm Hadronic}$        & 0.808 (3)(15)  &  0.805 (3)(17)  \\ 
$Z_q^{\rm Ward}$ & 0.753 (16)(30) &  0.750 (15)(30) \\ 
\end{tabular}
\caption{Final $Z$-factor results. $Z_q$ is calculated two ways: from
using $Z_A/Z_q$ from this table combined with $Z_A$ from hardonic matrix elements,
denoted $Z_q^{{\rm Hadronic}}$, and from the conserved currents using off-shell
quark states, denoted $Z_q^{{\rm Ward}}$.}
\label{tab}
\end{table}

\begin{table}[!htb]
\begin{tabular}{cccc} 
$(ap)^2$ & $1/\Lambda_A$ & $1/\Lambda_S$ & $1/\Lambda_T$ \\ 
\hline
0.501 & 0.9225(26) & 0.6200(86) & 1.0542(38) \\ 
0.616 & 0.9240(23) & 0.6401(67) & 1.0446(30) \\ 
0.655 & 0.9208(22) & 0.6481(64) & 1.0352(28) \\ 
0.771 & 0.9206(20) & 0.6574(55) & 1.0273(22) \\ 
0.810 & 0.9178(23) & 0.6614(59) & 1.0197(27) \\ 
0.925 & 0.9187(21) & 0.6755(53) & 1.0125(25) \\ 
0.964 & 0.9172(22) & 0.6784(54) & 1.0092(24) \\ 
1.079 & 0.9157(27) & 0.6910(62) & 0.9997(34) \\ 
1.118 & 0.9164(23) & 0.6935(52) & 0.9989(26) \\ 
1.234 & 0.9185(20) & 0.7007(45) & 0.9989(23) \\ 
1.272 & 0.9150(23) & 0.7059(51) & 0.9920(26) \\ 
1.388 & 0.9147(21) & 0.7102(45) & 0.9889(24) \\ 
1.426 & 0.9141(25) & 0.7112(45) & 0.9876(29) \\ 
1.542 & 0.9106(26) & 0.7168(48) & 0.9801(30) \\ 
1.581 & 0.9104(25) & 0.7202(48) & 0.9775(27) \\ 
1.735 & 0.9080(29) & 0.7257(49) & 0.9706(32) \\ 
1.851 & 0.9101(28) & 0.7334(45) & 0.9727(31) \\ 
1.889 & 0.9081(30) & 0.7339(42) & 0.9706(35) \\ 
\end{tabular}
\caption{This table collects together the raw data
used for the ``Bare'' data plotted in Figures 14,15 and 16 for
$0.5 < (ap)^2 < 2.0$.}
\label{tab1} 
\end{table}

\begin{table}[!htb]
\begin{tabular}{cc}
$\beta_i$ & Quenched Value \\ \hline
$\beta_0$ & 11    \\ 
$\beta_1$ & 102   \\ 
$\beta_2$ & 1428.5\\ 
\end{tabular}
\caption{$\beta_i$'s for the Quenched theory}
\label{tab:bet}
\end{table}

\begin{table}[!htb]
\begin{tabular}{cc}
Elements of $\gamma_q$ & Quenched Value \\ 
\hline 
$\gamma^{(0)}$ & 0       \\ 
$\gamma^{(1)}$ & 44.6667 \\ 
$\gamma^{(2)}$ & 1056.65 \\ 
\end{tabular}
\caption{Quenched $Z_q$ Anomalous Dimensions}
\label{tab:zqr}
\end{table}

\begin{table}[!htb]
\begin{tabular}{cc}
Elements of $\gamma_S$ & Quenched Value \\ 
\hline 
 $\gamma^{(0)}$ & -8       \\
 $\gamma^{(1)}$ & -134.667 \\
 $\gamma^{(2)}$ & -2498    \\
\end{tabular}
\caption{Quenched $Z_S$ Anomalous Dimensions}
\label{tab:zqs}
\end{table}

\begin{table}[!htb]
\begin{tabular}{cc}
Elements of $\gamma_T$ & Quenched Value \\ 
\hline 
$\gamma^{(0)}$ &2.66667 \\
\end{tabular}
\caption{Quenched $Z_T$ Anomalous Dimension}
\label{tab:zqt}
\end{table}

\begin{table}[!htb]
\begin{tabular}{cc}
 $Z^{(1)}_0$ & 0\\
 $Z^{(2)}_0$ &-14.4975\\
\end{tabular}
\caption{$Z_q$ RI to $\overline{MS}$ matching coefficients}
\label{app:matchq}
\end{table}

\begin{table}[!htb]
\begin{tabular}{cc}
 $Z^{(1)}_0$ &5.33333\\
 $Z^{(2)}_0$ &188.651\\
\end{tabular}
\caption{$Z_S$ RI to $\overline{MS}$ matching coefficients}
\label{app:matchS}
\end{table}

\newpage

\clearpage
\newpage

\clearpage
\newpage

\clearpage
\newpage

\clearpage
\newpage

\clearpage
\newpage

\clearpage
\newpage

\clearpage
\newpage

\clearpage
\newpage

\clearpage
\newpage

\clearpage
\newpage

\clearpage
\newpage

\clearpage
\newpage

\clearpage
\newpage

\clearpage
\newpage

\clearpage
\newpage

\clearpage
\newpage

\clearpage
\newpage

\clearpage
\newpage

\clearpage
\newpage

\clearpage
\newpage

\end{document}